\documentclass[aps,prl,10pt,twocolumn]{revtex4-2}
\usepackage{xcolor}
\usepackage{times}
\usepackage{changes}
\usepackage{ulem}
\usepackage[percent]{overpic}
\usepackage{graphicx}
\usepackage{amssymb}

\usepackage{epstopdf}
\usepackage{amsmath}

\usepackage{bbold}
\usepackage{enumitem}
\usepackage{bm}
\usepackage{physics}
\usepackage{braket}
\usepackage{ifthen}
\usepackage{picture}  

\makeatletter
\graphicspath{{Images/}}

\usepackage{hyperref}
\hypersetup{
	colorlinks=true,
	linkcolor=black,          
	citecolor=blue,           
	filecolor=magenta,        
	urlcolor=cyan
}

\makeatother

\newboolean{showskeleton}
\setboolean{showskeleton}{false} 

\newcommand{\skeleton}[1]{
    \ifthenelse{\boolean{showskeleton}}
    { {\color{blue!50!gray} #1} } 
    {}                     
}
\begin{document}

\title{Restoring thermalization in long-range quantum magnets with staggered magnetic fields}

\author{Lucas Winter}
\affiliation{Faculty of Physics, University of Vienna, Boltzmanngasse 5, 1090 Vienna, Austria}
\author{Pietro Brighi}
\affiliation{Faculty of Physics, University of Vienna, Boltzmanngasse 5, 1090 Vienna, Austria}
\author{Andreas Nunnenkamp}
\affiliation{Faculty of Physics, University of Vienna, Boltzmanngasse 5, 1090 Vienna, Austria}

\date{\today} 

\begin{abstract}
Quantum systems with strong long-range interactions are thought to resist thermalization because of their discrete energy spectra. We show that applying a staggered magnetic field to a strong long-range Heisenberg antiferromagnet restores thermalization for a large class of initial states by breaking  permutational symmetry. Using self-consistent mean-field theory and exact diagonalization, we reveal that the energy spectrum, while composed of discrete subspaces, collectively forms a dense spectrum. The equilibration time is independent of system size and depends only on the fluctuations in the initial state.
 For initial states at low to intermediate energies, the long-time average aligns with the microcanonical ensemble. However, for states in the middle of the spectrum the long-time average depends on the initial state due to quantum scar-like eigenstates localized at unstable points in classical phase space.
Our results can be readily tested on a range of experimental platforms, including Rydberg atoms or optical cavities.
\end{abstract}

\maketitle

 \textit{Introduction.---} A central challenge in non-equilibrium quantum physics is to understand the mechanisms by which isolated quantum systems approach thermal equilibrium. In many instances, the Eigenstate Thermalization Hypothesis (ETH) provides a successful framework for explaining how systems thermalize~\cite{deutsch_quantum_1991, srednicki_chaos_1994, dalessio_quantum_2016}. However, there are exceptions to ETH where thermalization in many-body systems fails, for example due to many-body localization~\cite{ gornyi_interacting_2005, basko_problem_2006, oganesyan_localization_2007, schreiber_observation_2015, smith_many-body_2016, abanin_colloquium_2019}, Hilbert-space fragmentation~\cite{khemani_localization_2020, sala_ergodicity-breaking_2020}, and integrability~\cite{takahashi_thermodynamics_1999, kinoshita_quantum_2006, rigol_relaxation_2007}.

Quantum systems with strong long-range (LR) interactions -- where the interaction strength decays as $1/r^\alpha$ with $\alpha < d$ ($d$ is the dimension)~\cite{defenu_out--equilibrium_2024, defenu_long-range_2023} -- have been shown to violate several principles that ensure thermalization in short-ranged systems, such as locality, ensemble equivalence, and additivity~\cite{barre_inequivalence_2001, thirring_systems_1970, mukamel_statistical_2008, defenu_ensemble_2024}.
In the literature, two main mechanisms have been put forward to explain why these systems fail to equilibrate.
First, strong long-range interactions lead to a discrete energy spectrum in the thermodynamic limit~\cite{last_quantum_1996, defenu_metastability_2021}, with energy gaps protecting metastable states whose lifetimes grow with system size~\cite{ campa_statistical_2009, gabrielli_quasistationary_2010, joyce_relaxation_2010}. 
Second, the permutational symmetry in the fully-connected limit partially persists for $0<\alpha<d$, inducing structure in the eigenstates, effectively violating ETH and hindering thermalization~\cite{sugimoto_eigenstate_2022, russomanno_quantum_2021, mattes_long-range_2025}.

 Previous work has predominantly focused on how the range of interactions $\alpha$ affects thermalization~\cite{defenu_metastability_2021, mattes_long-range_2025, sugimoto_eigenstate_2022, russomanno_quantum_2021, gabrielli_quasistationary_2010, joyce_relaxation_2010, bachelard_universal_2013, eisert_breakdown_2013, metivier_spreading_2014, rajabpour_quantum_2015}. More recently, it has been shown that reducing the symmetry of initial states can lead to long-time coherent oscillations~\cite{iemini_dynamics_2024} and super-exponential scrambling dynamics~\cite{qi_surprises_2023}. However, whether thermalization can be restored -- independent of interaction range $\alpha$ -- by reducing the symmetry at the level of the Hamiltonian remains an exciting  and experimentally relevant open question.

\begin{figure}
    \includegraphics[width=\linewidth]{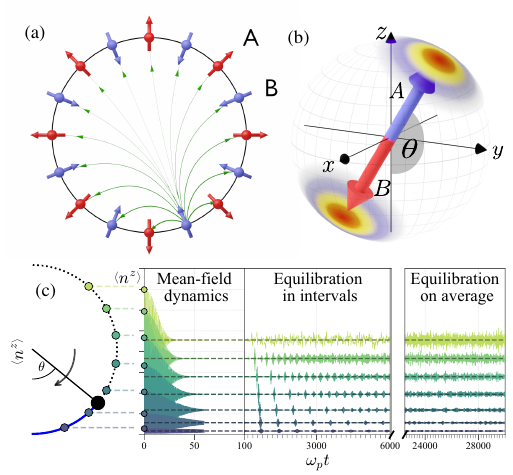}
    \caption{\textbf{Dynamics of a LR Heisenberg antiferromagnet.} (a) Illustration of the system \eqref{eq:hamiltonian}; a spin chain with two sublattices $A$ (red) and $B$ (blue) coupled via power-law interactions.
    (b) Bloch sphere representation of a mixed Néel state \eqref{eq:mixedneelstate}, parametrized by angle $\theta$ and uncertainty $\sigma$ in the orientation (increasing from red to blue).
    (c) Time evolution of the observable $\braket{n^z} = (\braket{S_A^z} - \braket{S_B^z})/N$, for mixed Néel states \eqref{eq:mixedneelstate}, features three distinct dynamical stages:
    mean-field pendulum-like oscillations at short times, periodic revivals, and eventual equilibration on average to the microcanonical ensemble (dashed lines). Results are obtained via exact diagonalization of Eq.~\eqref{eq:hamiltonianAlpha0} with $J/h=5$, $N=500$, $\alpha = 0$, and $\sigma = 10^{-2}$.}
\label{fig:one}
\end{figure}

In this letter, we demonstrate that a staggered magnetic field can restore thermalization in a long-range Heisenberg antiferromagnet for a large class of initial states by breaking full permutational symmetry.
Using self-consistent mean-field theory and exact diagonalization, we show the spectrum consists of $N^2$ discrete subspaces, where $N$ is the system size. 
Collectively, they form a dense spectrum. Equilibration progresses through three distinct dynamical stages: (i) oscillatory mean-field dynamics resembling classical pendulum motion, (ii) intermediate-time equilibration with periodic revivals, and (iii) long-time equilibration on average. In stark contrast with permutationally-invariant LR systems, the equilibration time is independent of system size and depends only on the fluctuations in the initial state.
For initial states at low to intermediate energies, the long-time average aligns with the microcanonical ensemble, indicating thermalization. However, for states close to the energy dividing bound and free trajectories in classical phase space, ergodicity breaks down, and the long-time average depends on the initial state. This behavior stems from quantum scar-like eigenstates localized at the unstable point.

\textit{The Model.---} We consider a long-range quantum Heisenberg antiferromagnet in a staggered magnetic field 
\begin{align}
H = \frac J {4Z_\alpha}\sum_{i,j } \frac{\boldsymbol \sigma_{i}\cdot \boldsymbol \sigma_{j}}{r_{ij}^\alpha}   + \frac h2  \sum_{j}\left( \sigma_{A,j}^z -  \sigma_{B,j}^z\right).\label{eq:hamiltonian}
\end{align}
Specifically, the geometry is that of a one-dimensional spin chain with sublattices $A$ and $B$ and periodic boundary conditions [Fig.~\ref{fig:one} (a)]. The operators $\boldsymbol \sigma_{A,i}$ and $\boldsymbol \sigma_{B,i}$ are spin-1/2 Pauli operators on site $i$ and obey standard commutation relations. The first term represents an antiferromagnetic exchange interaction ($J > 0$) between all spins. Here, $\boldsymbol\sigma_{i} = \boldsymbol \sigma_{A,i} + \boldsymbol \sigma_{B,i}$ is the composite spin at site $i$. Crucially, the interaction strength decays with distance as a power law, $1/r_{ij}^\alpha$. We consider the strong long-range interaction regime with $0 \le \alpha < 1$. The normalization factor $Z_\alpha = (1- \alpha) 2^{1- \alpha} N^{\alpha-1}$ ensures that the total energy is extensive ~\cite{kac_van_1963, defenu_out--equilibrium_2024}. The second term introduces a staggered magnetic field of strength $h$ along the $z$ direction. This field breaks the inherent symmetry between the $A$ and $B$ sublattices and the rotation symmetry of the Heisenberg interactions.

We investigate system dynamics starting from mixed Néel states, $\rho(\Omega_0, \sigma)$. These states represent different orientations, $\Omega_0 = (\theta_0, \phi_0)$, of the Néel vector defined as $n^z = \frac{1}{N}\sum_j (\sigma_{A, j}^z - \sigma_{B, j}^z)$. Crucially, this allows for tuning the fluctuations $\sigma_{nz}^2 = \langle (n^z)^2 \rangle - \langle n^z \rangle^2$  independent of system size [Fig.~\ref{fig:one} (b)]. Specifically, the mixed Néel state is a statistical ensemble of pure Néel states, $\ket{\uparrow}_A \otimes \ket{\downarrow}_B$, rotated by angles $\theta$ and $\phi$ according to $\ket{\theta, \phi} =   e^{i\phi S^z}e^{i\theta S^x}\ket{\uparrow}_A \otimes \ket{\downarrow}_B$. Here, $S^\gamma = \frac 12 \sum_j(\sigma_{A, j}^\gamma + \sigma_{B, j}^\gamma)$ is the collective spin operator.
In the mixed Néel state, these pure states are weighted by a Gaussian factor according to 
\begin{align}
\rho(\Omega_0, \sigma) = \frac{1}{Z} \int d\Omega \, \exp\bigg({-\frac{(1 - \boldsymbol r(\Omega)\cdot \boldsymbol r(\Omega_0) )^2}{2\sigma^2}}\bigg)\  \ket{\Omega} \bra{\Omega}.
\label{eq:mixedneelstate}
\end{align}
Here, $ Z $ is the normalization constant, and $ \boldsymbol{r}(\Omega)$ is the unit vector in spherical coordinates. We note that even for $\sigma\to 0$, the mixed Néel state still exhibits quantum fluctuations $\sigma_{nz}^2 \propto 1 /N$ (see Supplemental Material).

\textit{Dynamics.---} Let us first focus on the fully connected limit ($\alpha =0$). This limit shares many of the qualitative features with the finite $\alpha>0$ case but crucially allows for comparison of analytical results with large-scale numerical diagonalization. The Hamiltonian depends on the collective spin operators $\boldsymbol S_{A/B} = \frac 12 \sum_{j} \boldsymbol \sigma_{A/B,j}$. To further simplify the analysis, we introduce the total spin $\boldsymbol S = \boldsymbol S_A + \boldsymbol S_B$ and the staggered spin $\boldsymbol N = \boldsymbol S_A - \boldsymbol S_B$. The Hamiltonian \eqref{eq:hamiltonian} reduces to  
\begin{align}
        H = \frac JN\boldsymbol S^2 + h N^z \ .\label{eq:hamiltonianAlpha0}
\end{align}
Using the well-established Schwinger boson mapping the Hamiltonian \eqref{eq:hamiltonianAlpha0} can be diagonalized for systems up to hundreds of spins (see Supplemental Material).

 For short times, the mean-field equations of motion are expected to accurately describe the dynamics of long-ranged interacting systems. In our system, when the total spin $\boldsymbol{S}^2$ is small, the angle $\theta$ of the Néel parameter behaves like a simple pendulum, $\ddot{\theta}  +\omega_p^2 \sin(\theta)=0$ ~\cite{kimel_inertia-driven_2009}, where $\omega_p = \sqrt{Jh}$. At low energies, the system exhibits bound oscillations, while at high energies, it transitions to free rotations. These regimes are divided by a separatrix at energy $E_s$, at which the pendulum  approaches the unstable point $\theta=\pi$ exponentially slowly.
 
We compare this mean-field dynamics to the exact quantum evolution of mixed Néel states \eqref{eq:mixedneelstate} for the Hamiltonian \eqref{eq:hamiltonianAlpha0} [Fig.~1(c)]. Initially, $\braket{n^z}$ follows the simple pendulum mean-field dynamics. At longer times, however, the pendulum-like oscillations decay, give way to a complex series of revivals, and ultimately equilibrate to the microcanonical average.

\begin{figure}
    \centering
    \includegraphics[width=\linewidth]{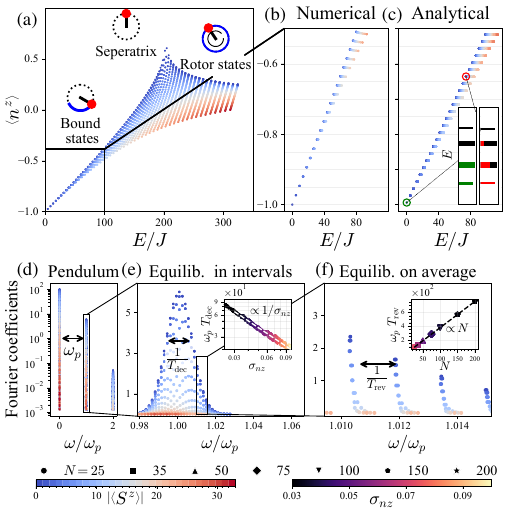}
    \caption{\textbf{Eigenspectrum and Néel vector dynamics.}
    (a) Eigenstate expectation value $\braket{n^z}$ as a function of energy $E/J$, color-coded by $\langle S^z\rangle$ (blue to red) according to exact diagonalization of Eq.~\eqref{eq:hamiltonianAlpha0}. Bound, separatrix, and rotor states reflect classical trajectories in phase space.
    Close to the ground state, numerical results (b) agree well with analytical many-body energies (c).
    The inset of (c) shows how single-particle levels are occupied to form representative many-body eigenstates.
    Figure~(d)--(f). Fourier analysis of $\langle n^z(t)\rangle$ mirroring the real-time dynamics in Fig.~1(c).
    (d) A dominant peak at $\omega_p$ corresponds to mean-field pendulum oscillations.
    (e) Gaussian broadening of each peak induces periodic decay and revivals, with the decay time $T_{\mathrm{dec}}\propto 1/\sigma_{nz}$ (inset).
    (f) Closer inspection reveals equally spaced level splittings setting the revival period $T_{\mathrm{rev}}\propto N$ (inset); a smaller quadratic splitting leads to equilibration on average. The system parameters are $J/h = 2, N=50$, and $\alpha =0$ and the initial state is a mixed Néel state \eqref{eq:mixedneelstate} with $\theta_0 = \pi/4$ and $\sigma = 10^{-2}$.}
\end{figure}

\skeleton{\noindent \textbf{Paragraph 7: Analytical theory} \\ 
\noindent \textit{\textbf{Message: } There are $N^2$ discrete subspaces forming a dense spectrum.}\\ }

\textit{Spectrum.---} To understand the long-time dynamics, we examine the spectrum. Fig.~2(a) shows the expectation value $\langle n^z \rangle$ in eigenstates as a function of their energy. At low energies, the mean-field equation predicts bound pendulum oscillations. In this regime, the spectrum is single-valued. With increasing energy, the dependence of $\langle n^z \rangle$ on the magnetization sector $S^z$ (red to blue) increases reaching a maximum at the separatrix energy $E_s$. At very high energy, there are eigenstates corresponding to free pendulum rotations.

We will now analytically diagonalize the Hamiltonian in the region corresponding to bound oscillations. Similar to the approach used for permutationally-invariant LR systems\  in Ref.~\cite{defenu_metastability_2021}, we apply the Jordan-Wigner transformation to map the spin operators to fermions. A detailed step-by-step derivation is presented in the Supplemental Material. Crucially, the long-ranged nature of the interactions allows us to replace string operators with their mean-field expectation values. In contrast to the permutationally-invariant \  case, where the string operator expectation values are fixed to one~\cite{defenu_metastability_2021}, here they depend on the expectation values of the collective operators $\braket{s^z} = \braket{S^z}/N$ and $\braket{n^z} = \braket{N^z}/N$. To evaluate these string operator expectation values, we make three key assumptions. (i) We apply a mean-field decoupling $\braket{S_A^z S_B^z} = \braket{S^z}^2 - \braket{N^z}^2$. (ii) We consider the spectrum close to the ground state where $n^z\approx-1$ and $s^z\approx 0$. (iii) We assume permutational invariance of the two-point spin correlation function $\braket{\sigma_{i, L}^z\sigma_{j,M}^z} =\braket{\sigma_{i',L}^z \sigma_{j', M}^z}$, where sites $i$ and $i'$ belong to the same sublattice ($L$), and sites $j$ and $j'$ belong to the same sublattice ($M$). This reflects the bipartite nature of the system. After obtaining a quadratic Hamiltonian, we transform to Fourier space, 
\begin{equation}
H =-2 \sum_{n} \begin{pmatrix} c_{A, n}^\dagger \\ c_{ B, n}^\dagger \end{pmatrix}
\begin{pmatrix}
h_{N, n} & \Gamma_n \\
\Gamma_n^* & - h_{N, n}
\end{pmatrix}
\begin{pmatrix} c_{A, n}^{\phantom{\dagger}} \\ c_{B, n}^{\phantom{\dagger}}  \end{pmatrix} + JN\braket{s^z}^2
\label{eq:mean_field_Hamiltonian}
\end{equation}

\noindent where $c_{A/B,n}$ are fermionic operators with standard anticommutation relations. The coefficients $h_{N,n}$ and $\Gamma_n$ depend on the collective variables $\langle s^z\rangle$ and $\langle n^z\rangle$ (see Supplemental Material) and on the form factor
\begin{equation}
    F_\alpha(n)  = (1-\alpha)2^{1- \alpha}\int_0^{1/2}ds\ \frac{\cos(2\pi s n)}{s^\alpha} \ . \label{eq:formfactor}
\end{equation}

\noindent Unlike in short-ranged systems, where the form factor is typically a smooth function of a continuous wave vector $k$, here $F_\alpha(n)$ depends on a discrete index $n$. This means even in the thermodynamic limit, $|F_{\alpha}(n+1)|- |F_{\alpha}(n)|$ is finite.

Diagonalizing the Hamiltonian \eqref{eq:mean_field_Hamiltonian} reveals an energy spectrum composed of $N^2$ discrete subspaces. This arises from the energy levels $E_{A/B, n} =  \pm \sqrt{h_{N, n}^2 + |\Gamma_n|^2} + JN\braket{s^z}^2$ [inset of Figure 2(c)] inheriting the properties of the discrete form factor \eqref{eq:formfactor}. Thus, the level spacings in each subspace stay finite $E_{A/B,n+1} - E_{A/B, n}>0$, even in the thermodynamic limit $N\to\infty$. The  energy also depends on the magnetization sector $\braket{s^z}$ and Néel parameter  $\braket{n^z}$, with each of the $N^2$ pairings defining a unique subspace. Note, here $\braket{n^z}$ labels the subspaces, but it is not a quantum number as $N^z$ does generally not commute with $H$.
Crucially, the number of subspaces $\mathcal O(N^2)$ scales faster than in permutationally-invariant LR systems\ $\mathcal O(N)$. This is a consequence of the staggered magnetic field reducing full permutational symmetry to bipartite symmetry and therefore lifting the degeneracy in $\braket{n^z}$.

Although each individual subspace is discrete, the scaling with the number of subspaces leads to a dense spectrum. As the system is extensive, the energy range scales as $\mathcal O(N)$. However, there are $N^2$ subspaces meaning the average energy spacing must scale as $\mathcal O(N)/N^2 = \mathcal O(1/N)$. Therefore, in the thermodynamic limit $N\to \infty$, the average energy spacing vanishes and the spectrum becomes dense. This contrasts sharply with permutationally-invariant LR systems, where only $N$ subspaces exist, meaning the average level spacing $\mathcal O(N)/N = \mathcal O(1)$ can remain finite~\cite{defenu_metastability_2021,lerose_theory_2025} and the total spectrum is discrete.  Note, this argument generally implies that any long-range system in which breaking of permutational symmetry produces at least $N^2$ non-degenerate energy levels will exhibit a dense spectrum, provided these levels remain non-degenerate in the thermodynamic limit.

\textit{Equilibration.---}  The organization of the energy spectrum into discrete subspaces, which collectively form a dense spectrum, shapes the equilibration dynamics. Intuitively, states exhibiting finite fluctuations $\sigma>0$ equilibrate, as they span multiple  subspaces collectively forming a dense spectrum.

Using the analytical spectrum we can now understand the dynamics of $\braket{N^z}(t)$ in the case $\alpha = 0$ in Fig.~1 (c). Motivated by numerical results, we assume (i) $N^z$ is predominantly diagonal in the energy eigenbasis, with non-zero matrix elements for transitions to adjacent eigenstates $N^z \to N^z \pm 1$, (ii) we approximate the mixed Néel states (\ref{eq:mixedneelstate}) as a superposition of energy eigenstates $|s^z, n^z, j\rangle$ with Gaussian coefficients. The $n^z$ distribution is centered at $n_0^z = \cos(\theta_0)$ with width $\sigma_{nz}$, while the $s^z$ distribution is centered at zero with width $\sigma_{sz}$.
Assuming the fluctuations of the initial state are small $\sigma_{nz}, \sigma_{sz} \ll 1$, we obtain
\begin{align}
\braket{N^z}(t) = N_{\mathrm{stat}}^z + \mathcal N^z_{\mathrm{pend}}(t) \mathcal N^z_{\mathrm{comb}}(t) \mathcal N^z_{\mathrm{eq}}(t)\ .
\end{align}

Each term corresponds to a behavior at different timescales, manifesting as progressively finer structures in the Fourier spectrum [Figs.~2(d)-(f)]. At the shortest timescales, a prominent peak at frequency $\omega_p$ [Fig.~2(d)] corresponds to pendulum-like oscillations $\mathcal N^z_{\mathrm{pend}}(t) = (N_0^z - N_{\mathrm{stat}}^z)\cos(\omega_p t)$ [Fig.~1(c)].  Here the dynamics match the mean-field equations discussed at the beginning. At intermediate timescales, this peak splits into a Gaussian frequency comb [Fig.~2(e)]. This leads to periodic decay and revival of pendulum oscillations, described by $\mathcal N_{\mathrm{comb}}^z(t) =\sum_n \exp\left(-(t - nT_{\mathrm{rev}})^2/{T_{\mathrm{dec}}^2}\right)$.
The decay time, $T_{\mathrm{dec}}$, is determined by the frequency comb's width [Fig.~2(e)]. It is inversely proportional to initial-state fluctuations $T_{\mathrm{dec}} \propto 1/(\sigma_{nz})$ but importantly not system size $N$, as confirmed by numerical simulations [Fig.~2(e) inset]. Conversely, the revival time, $T_{\mathrm{rev}}$, depends on the frequency comb's spacing [Fig.~2(f)] scaling with system size $T_{\mathrm{rev}}\propto N$ [Fig.~2(f) inset]. 
Finally, over extended timescales the finest structures in the spectrum are  resolved and equilibration is characterized by $\mathcal{N}^z_{\mathrm{eq}}(t)$. Antiferromagnetic fluctuations impose a quadratic frequency spacing $[\propto (s^z)^2]$, evident in Figure 2(f) (bottom). This breaks the regularity of the frequency comb, causing the oscillation revivals to decay according to the envelope $|\mathcal{N}^z_{\mathrm{eq}}(t)|\propto 1/\sqrt[4]{1 + ( t/{T_{\mathrm{eq}}})^2}$. Again, the equilibration time is inversely proportional to fluctuations in the initial state $T_{\mathrm{eq}} \propto 1/\sigma_{sz}^2$.

The analytical and numerical results confirm that the equilibration timescales of $N^z$ ($T_{\text{dec}} \propto 1/\sigma_{nz}$, and $T_{\mathrm{eq}}\propto 1/\sigma_{sz}^2$) depend on the fluctuations in the initial state. For a state with finite fluctuations, the system always equilibrates in finite time. In the case $\sigma \to 0$, instead, the equilibration timescales diverge in the thermodynamic limit, $T_{\mathrm{dec}}\propto \sqrt{N}$ and $T_{\mathrm{eq}}\propto N$ (as quantum fluctuations vanish with $\sigma_{nz}^2, \sigma_{sz}^2 \propto 1/N$). While this resembles metastability in other strong LR systems~\cite{defenu_metastability_2021, davis_protecting_2020}, the effect here hinges on diminishing fluctuations. The equilibration time does not explicitly scale with system size, implying that the dynamics are not protected by energy gaps in the spectrum.
We note that the reduced density matrix $\mathrm{Tr}_{S^z}(\rho)$ also equilibrates, in the time $T_{\mathrm{eq}}$, to the diagonal ensemble determining the long-time averages. Crucially, this generalizes equilibration to all observables commuting with $S^z$ (see Supplemental Material).

For finite-range interactions $\alpha > 0$, a new timescale $T_{\mathrm{spin}}$ emerges controlling the decay of the collective spins, $\langle \boldsymbol{S}_A^2\rangle$ and $\langle \boldsymbol{S}_B^2\rangle$ (see Supplemental Material). If $\alpha =0$, all $N^2$ subspaces are fully degenerate. For $\alpha >0$, the energy levels of each subspace split, and the width of the resulting multiplets determines  $T_{\mathrm{spin}}$. However, as system size increases, the energy levels within each subspace accumulate at a single point, effectively reducing the multiplet's width.
This implies the lifetime of the spin lengths increases with system size, contrasting with the Néel parameter where it depends on the fluctuations. Consequently, in the thermodynamic limit, the lifetime of the collective spins $\braket{\boldsymbol S_{A/B}^2}$ diverges, so that the dynamics for systems with $0 < \alpha < 1$ approximately match those of systems with $\alpha = 0$. \\

\begin{figure}
\includegraphics[width =\linewidth]{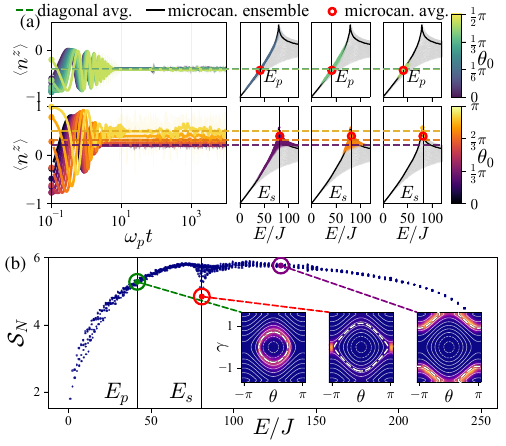}
\caption{\textbf{Breakdown of ergodicity at the separatrix.} (a) Left: Long-time evolution of $\braket{n^z}$
for initial states at energy $E_0$ with varying initial angles $\theta_0$ (different shading). Right: Energy spectrum; eigenstates are colored by their overlap with the corresponding initial state on the left. 
 Black line: microcanonical average versus energy. Red circles: microcanonical averages of initial states. Dashed lines: diagonal averages of initial states.
Top row $(E_0 =E_p< E_s$): The long-term average of $\braket{n^z}$ is independent of $\theta_0$.
Both ensembles agree. Bottom row ($E_0= E_s$): The long-term average of $\braket{n^z}$ clearly shows a dependence on $\theta_0$
and the diagonal ensemble average deviates from the microcanonical average.  (b) Entanglement entropy $\mathcal{S}_N$
of energy eigenstates for the $A-B$ partition for eigenstates   with $\braket{\boldsymbol S^2}/N^2 < 0.2$. Insets: Husimi-$Q$ distributions for selected eigenstates, white dashed line shows accessible phase space at that eigenenergy. Near the separatrix, eigenstates exhibit low $\mathcal S_N$ and are localized in an unstable phase space region. System parameters $J/h=5, N=50$, and $\alpha =0$.
}
\end{figure}

\textit{Ergodicity of observables.---} Thermalization demands not only the equilibration of observables but also that  long-time averages match those in the microcanonical ensemble~\cite{gogolin_equilibration_2016}. We study the evolution of initial states with varying angles $\theta_0$ at fixed energy $E_0$. To achieve this, in the pendulum analogy, finite initial momentum is required. For the spin system, this corresponds to tilting the spins $\boldsymbol{S}_A$ and $\boldsymbol{S}_B$ by an angle $\gamma_0$, yielding the state  
$\ket{\theta_0,\gamma_0} = e^{i\theta_0 S^x} \left( e^{i\gamma_0 S^x} \ket{\downarrow}_A \otimes e^{-i\gamma_0 S^x} \ket{\uparrow}_B \right).$
Here, $\gamma_0 = 0$ recovers the Néel state.  Note at any finite system size this state equilibrates due to its finite quantum fluctuations. 
We observe two distinct dynamical regimes: (i) Away from the separatrix $E_0 = E_p < E_s$, all trajectories converge to the microcanonical average [Fig.~3(a)] and (ii) at the separatrix energy $E_0 = E_s$, the long-time averages depend on the initial angle and deviate from microcanonical ensemble [Fig.~3(b)], indicating ergodicity is broken.

The discrepancy at the separatrix stems from the broad range of $\langle n^z\rangle$ and from the uneven distribution of the initial states across the eigenstates.
These determine the long-time average through the diagonal ensemble, which weights each eigenstate $\ket j$ expectation value $\braket{j|n^z|j}$ by the overlap with the initial state $|\braket{j|\theta_0, \gamma_0}|^2$.
Conversely, microcanonical averaging gives equal weight to all eigenstates within a narrow energy window. A discrepancy between ensembles occurs when a small subset of eigenstates exhibits an anomalously large overlap with $\lvert \theta_0, \gamma_0\rangle$  and the distribution of expectation values at that energy is wide. 
Away from the separatrix [Fig.~3(a) top right], the overlap distribution is roughly homogeneous. However, at the separatrix [Fig.~3(a) bottom right] rare high-overlap eigenstates dominate the diagonal ensemble and hence determine the long-time average.

We further investigate the ergodicity breaking eigenstates at the separatrix energy $E_s$. Analyzing the entanglement entropy, $\mathcal{S}_N = -\mathrm{Tr}\  \rho_A \ln \rho_A$, where $\rho_A = \mathrm{Tr}_B |\psi\rangle\langle\psi|$, reveals a significant reduction for these eigenstates [see Fig.~3(b)].  While this is reminiscent of the reduced entanglement entropy in quantum many-body scars \cite{turner_quantum_2018}, here the ergodicity breaking does not manifest as persistent oscillations but as non-thermal long-time averages.
To explore the structure of these eigenstates, we examine the Husimi $Q$-distribution, $Q(\theta, \gamma) = |\langle \theta, \gamma | \psi \rangle|^2$, over the classical pendulum phase space $(\theta, \gamma)$ [Fig.~3(b), inset]. Away from the separatrix, eigenstates spread across the entire phase-space segment accessible at that energy (white dashed line). At the separatrix, however, they localize in a small region near the unstable fixed point $\theta = \pi$, corresponding to the upright pendulum. This can be understood from the classical dynamics, as the Néel parameter exponentially slows down at the unstable point. This phenomenon is reminiscent of quantum single-particle scars~\cite{bogomolny_smoothed_1988, berry_quantum_1997, hudomal_quantum_2020, evrard_quantum_2024}. However, in this case the eigenfunctions are predominantly localized at a single unstable point rather than along an unstable periodic orbit.

{\textit{Conclusion.---}We showed that a staggered magnetic field can restore equilibration in a strong long-range antiferromagnet for initial states with finite fluctuations. The magnetic field breaks permutational symmetry leading to a dense spectrum in the thermodynamic limit.
While the long-time averages of low-energy initial states align with those in the microcanonical ensemble, they become initial-state dependent in the middle of the spectrum, reflecting quantum scar-like eigenstates arising from an unstable classical phase-space point.
}

 We presented a minimal example demonstrating that thermalization can be restored for collective observables independently of the interaction range. Note that the staggered magnetic field leaves permutation symmetry within each sublattice conserved implying that thermalization cannot be expected for arbitrary operators in this setting. This invites further research into the role of symmetries in the dynamics of LR systems beyond the bipartite symmetry considered here.

In addition to long-ranged Heisenberg interactions, our proposal requires only local staggered magnetic fields, making it readily implementable in existing experimental platforms such as trapped ions~\cite{blatt_quantum_2012, richerme_non-local_2014, kranzl_observation_2023, garttner_measuring_2017}, Rydberg atom arrays~\cite{steinert_spatially_2023, zeiher_coherent_2017}, and atoms coupled to cavities~\cite{mivehvar_cavity_2021, davis_protecting_2020}.

\textit{Acknowledgments.---}
This research was funded in whole or in part by the Austrian Science Fund (FWF) [10.55776/COE1, 10.55776/ESP9057324].
For Open Access purposes, the authors have applied a CC-BY public copyright license to any author accepted manuscript version arising from this submission.

\bibliography{references} 

\end{document}


\title{Supplemental Material for "Restoring thermalization in long-range quantum magnets with staggered magnetic fields"}

\author{Lucas Winter}
\affiliation{Faculty of Physics, University of Vienna, Boltzmanngasse 5, 1090 Vienna, Austria}
\author{Pietro Brighi}
\affiliation{Faculty of Physics, University of Vienna, Boltzmanngasse 5, 1090 Vienna, Austria}
\author{Andreas Nunnenkamp}
\affiliation{Faculty of Physics, University of Vienna, Boltzmanngasse 5, 1090 Vienna, Austria}

\maketitle

\tableofcontents

\section{\texorpdfstring{Schwinger-Boson mapping in the case $\alpha = 0$}{Schwinger-Boson mapping in the case alpha = 0}}

\noindent In the main text we show in the case $\alpha = 0$ the Hamiltonian can be fully written in terms of collective spin operators $S_{L}^\gamma = \sum_j \sigma_{L, j}^\gamma$ with $\gamma \in \{x,y,z\}$ and $L \in \{A,B\}$

\begin{align}
 H = \frac J  N \mathbf{S}_A \cdot \mathbf{S}_B + h (S_A^z - S_B^z). \end{align} 
 
\noindent  Below we outline the application of the Schwinger boson representation to the collective spin Hamiltonian. This will enable exact diagonalization on finite systems up to large system sizes.

The Schwinger boson representation provides a convenient way to represent spin operators using two species of bosons. For each collective spin, we introduce two bosonic annihilation (and corresponding creation) operators $a_{L}, a_{L}^\dagger$ and $b_{L}, b_{L}^\dagger$. These operators satisfy the usual bosonic commutation relations
\begin{align}
[a_{L}, a_{L}^\dagger] = 1, \quad [b_{L}, b_{L}^\dagger] = 1,
\end{align}

\noindent with all other commutators being zero.

The collective spin operators $S_{L}^{L}$ can be represented as 
\begin{align}
S_{L}^z = \tfrac{1}{2}(a_{L}^\dagger a_{L} - b_{L}^\dagger b_{L}), 
\qquad S_{L}^+ = a_{L}^\dagger b_{L}, 
\qquad S_{L}^- = b_{L}^\dagger a_{L}.
\end{align}
Here, $S_{L} = N_{L}/2$ where $N_{L}$ is the number of spin-$1/2$ constituents in ensemble ${L}$. The physical subspace is restricted by the constraint
\begin{equation}
a_{L}^\dagger a_{L} + b_{L}^\dagger b_{L} = N_{L}.
\end{equation}
This ensures that the dimension of the Hilbert space matches that of a spin-$S_{L}$ space.

We first express the isotropic interaction $\mathbf{S}_A \cdot \mathbf{S}_B$ in terms of bosons. Decompose it as
\[
\mathbf{S}_A \cdot \mathbf{S}_B = S_A^z S_B^z + \tfrac{1}{2}(S_A^+ S_B^- + S_A^- S_B^+).
\]
Substituting the Schwinger boson expressions, we have
\begin{align}
S_A^z S_B^z &= \tfrac{1}{4}(a_A^\dagger a_A - b_A^\dagger b_A)(a_B^\dagger a_B - b_B^\dagger b_B), \\
S_A^+ S_B^- &= (a_A^\dagger b_A)(b_B^\dagger a_B), \\
S_A^- S_B^+ &= (b_A^\dagger a_A)(a_B^\dagger b_B).
\end{align}
Thus,
\begin{equation}
\mathbf{S}_A \cdot \mathbf{S}_B = \tfrac{1}{4}(a_A^\dagger a_A - b_A^\dagger b_A)(a_B^\dagger a_B - b_B^\dagger b_B) 
+ \tfrac{1}{2}(a_A^\dagger b_A b_B^\dagger a_B + b_A^\dagger a_A a_B^\dagger b_B).
\end{equation}

\noindent The linear terms are straightforward. We have

\begin{align}
h (S_A^z - S_B^z) = \tfrac{h}{2} \bigl[(a_A^\dagger a_A - b_A^\dagger b_A) - (a_B^\dagger a_B - b_B^\dagger b_B)\bigr].
\end{align}

\noindent Collecting all terms, the Hamiltonian in the Schwinger boson representation is
\begin{align}
H &= \frac JN \biggl[ \tfrac{1}{4}(a_A^\dagger a_A - b_A^\dagger b_A)(a_B^\dagger a_B - b_B^\dagger b_B) 
+ \tfrac{1}{2}(a_A^\dagger b_A b_B^\dagger a_B + b_A^\dagger a_A a_B^\dagger b_B) \biggr] \nonumber \\
& \quad + \tfrac{h}{2} \bigl[(a_A^\dagger a_A - b_A^\dagger b_A) - (a_B^\dagger a_B - b_B^\dagger b_B)\bigr].
\end{align}

\noindent Importantly, the total magnetization along $z$,
\begin{equation}
S^z = S_A^z + S_B^z = \tfrac{1}{2}(a_A^\dagger a_A - b_A^\dagger b_A + a_B^\dagger a_B - b_B^\dagger b_B),
\end{equation}
commutes with the Hamiltonian $[H, S^z]=0$.  Thus, $S^z$ is a good quantum number. This conservation allows one to block-diagonalize the Hamiltonian according to different $S^z$ sectors, greatly simplifying the numerical effort in an exact diagonalization procedure.

When performing exact diagonalization, one works in a truncated bosonic Fock space that respects the constraints 
$a_{L}^\dagger a_{L} + b_{L}^\dagger b_{L} = N_{L}$
and uses the conservation of $S^z$ to reduce the size of the Hilbert space. 

In summary, the Schwinger boson representation transforms the spin Hamiltonian into a bosonic one, enabling the use of powerful numerical diagonalization techniques. The exact conservation of $S^z$ ensures that one can exploit symmetry to improve computational efficiency.

\section{Many-body spectrum of the Hamiltonian}
\label{sec:spectrum}
Below we present a detailed derivation and discussion of the long-range Heisenberg model [Eq. (1) of the main text] with a staggered field, providing further details on the steps outlined in the main text. We begin by introducing the model, specifying the notation, and carefully applying the Jordan-Wigner transformation to map spins to fermions. We then discuss how to handle the long-range interactions and the staggered field, and provide a consistent mean-field approximation for the string operators.

We consider a one-dimensional bipartite lattice composed of two sublattices, $A$ and $B$, each containing $N_L = N/2$ sites, with periodic boundary conditions. We label the lattice sites $j = 0, 1, \dots, N_L-1$ on sublattice $A$, and similarly $j=0,1,\dots, N_L-1$ on sublattice $B$. Thus, the total number of spins is $N = 2N_L$.

We define spin operators $\boldsymbol{\sigma}_{A,j} = (\sigma_{A,j}^x,\sigma_{A,j}^y,\sigma_{A,j}^z)$ on sublattice $A$ and similarly $\boldsymbol{\sigma}_{B,j}$ on sublattice $B$. The Hamiltonian we focus on is a long-range Heisenberg model with a staggered field (equivalent to Eq.~(1) of the main text)
\begin{align}
\boxed{H = \frac J{4Z_{\alpha}} \sum_{i, j}  \frac 1 {r_{ij}^\alpha} (\boldsymbol \sigma_{A,i}+ \boldsymbol{\sigma}_{B,i}) \cdot (\boldsymbol \sigma_{A,j}+ \boldsymbol{\sigma}_{B,j}) +\; \frac h 2\sum_{j}\left(\sigma_{A,j}^z - \sigma_{B,j}^z\right).}
\end{align}

\noindent Here, $r_{ij} = \min(|i-j|, N - |i-j])$ denotes the distance between spins at sites $i$ and $j$, and $\alpha$ controls the decay of the interaction strength. For simplicity, we often write terms symbolically as sums over $A$ and $B$ sublattices, assuming appropriate definitions of $r_{ij}$. The sign and factors are chosen such that $J>0$ corresponds to antiferromagnetic-like couplings between each pair of spins, and $h$ is a uniform staggered field, acting with $+h$ on $A$ sites and $-h$ on $B$ sites. Here $N^\alpha = \sum_{r= 1}^{N/2} r^{-\alpha}$   is a normalization factor ensuring the extensivity of the Hamiltonian. We have approximately 

\begin{align} Z_\alpha^{-1}\approx  (1- \alpha) 2^{1- \alpha} N^{\alpha-1} \label{eq:normalization}\ ,\end{align}

\noindent for $\alpha < 1$~\cite{defenu_metastability_2021}.

As explained in the main text, to analyze the problem in a fermionic language, we apply the Jordan-Wigner transformation. We introduce fermionic operators $c_{A,j}, c_{A,j}^\dagger$ and $c_{B,j}, c_{B,j}^\dagger$ which fulfill canonical anticommutation relations
\begin{align}
\{c_{\ell,L}, c_{\ell',L'}^\dagger\} = \delta_{\ell\ell'}\delta_{L,L'}, \quad L,L' \in \{A,B\}.
\end{align}

\noindent The Jordan-Wigner transformation maps spins to fermions as follows

\begin{align}
\sigma_{L,j}^x &= T_{j}^L(c_{L,j} + c_{L,j}^\dagger), \\
\sigma_{L,j}^y &= i T_{j}^L(c_{L,j} - c_{L,j}^\dagger), \\
\sigma_{L,j}^z &= 1 - 2c_{L,j}^\dagger c_{L,j}.
\end{align}

\noindent Here we have defined the string operators 

\begin{align}
    T_j^A &= \prod_{j<m} \sigma_{A,m}^z\sigma_{B,m}^z=\prod_{m<j} (1 - 2c_{A,m}^\dagger c_{A,m})(1 - 2c_{B,m}^\dagger c_{B,m})\ , \\ T_j^B &= T_j^A \sigma_{A,j}^z = T_j^A (1 - 2c_{A,j}^\dagger c_{A,j}) \ .
\end{align}

\noindent The string operators ensure the correct exchange statistics. Applying the string operators to the magnetic field term simply yields 

\begin{align}
    H_{\mathrm{mag}} = -2h\sum_{i} \bigg( c_{A,i}^\dagger c_{A,i} -  c_{B,i}^\dagger c_{B,i}\bigg) 
\end{align}

\noindent The Heisenberg interaction can be written as
\begin{align}
\boldsymbol{\sigma}_{L,i}\cdot \boldsymbol{\sigma}_{L',j} = \sigma_{L,i}^x \sigma_{L',j}^x + \sigma_{L,i}^y \sigma_{L',j}^y + \sigma_{L,i}^z \sigma_{L',j}^z.
\end{align}

\noindent We start by focusing on the $xy$ part of the interaction. The goal is to write this as an effective hopping interaction. We have 

\begin{align}
    \sigma_{L,i} ^x\sigma_{L',j}^x + \sigma_{L,i}^y \sigma_{L',j}^y = T_i^L T_j^{L'} \bigg( (c_{L,i}^\dagger  + c_{L,i}) ( c_{L',j}^\dagger +c_{L',j})) - (c_{L,i}  -c_{L,i}^\dagger)(c_{L',j}-c_{L',j}^\dagger) \bigg) = 2T_{ij}^{LL'}(c_{L,i}^\dagger c_{L',j } + c_{L,i}c_{L',j}^\dagger)  
\end{align}

\noindent where we have defined 

\begin{align}
    T_{ij}^{LL'} = \begin{cases} 
    \prod_{m = i+1}^{j-1}  \sigma_{A,m}^z \sigma_{B,m}^z \qquad &\mathrm{if} \ L = A, L'=B \\ \prod_{m=i+1}^{j-1} \sigma_{B,m}^z \sigma_{A,m}^z  \qquad &\mathrm{if}\ L = B, L'=  A \\ \sigma_{i+1, B}^z \prod_{m = i+2}^{j-1}  \sigma_{A,m}^z \sigma_{B,m}^z \qquad &\mathrm{if} \ L = A, L'=A \\ \sigma_{i+1, A}^z\prod_{m=i+2}^{j-1} \sigma_{B,m}^z \sigma_{A,m}^z  \qquad &\mathrm{if}\ L = B, L'=  B \ .  
    \end{cases} 
\end{align}

We next focus on the $zz$ part of the interaction. In this case we find 

\begin{align}
    \sum_{i,j} \sigma_{L,i}^z \sigma_{L',j}^z = (1 - 2c_{L,i}^\dagger c_{L,i})( 1 - 2c^\dagger_{L',j} c_{L',j}^{\phantom{\dagger}}) \ . 
\end{align}

\noindent This means the total Hamiltonian becomes with the Jordan-Wigner transformation 

\begin{align}
    H &= \frac J {4Z_{\alpha}}\sum_{i,j} \sum_{L, L' \in \{A,B\}} \frac 1 {r_{ij}}\bigg[ 2T^{LL'}_{ij} (c_{L,i}^\dagger c_{L',j} + \mathrm{h.c.} )  + (1-2c_{L,i}^\dagger c_{L,i}) (1-2c_{L',j}^\dagger c_{L',j})\bigg] \\ &-h\sum_{i} \bigg( c_{A,i}^\dagger c_{A,i} -  c_{B,i}^\dagger c_{B,i}\bigg)\ ,  
\end{align}

\noindent where we have neglected constant contributions to the Hamiltonian. We note the Hamiltonian contains highly non-local pairing of string operators. However, as local fluctuations are suppressed by long-range interactions we expect we can treat these terms accurately within a mean field approximation.

\subsection{Mean-field approximation}

Below, we provide further details on the mean-field approximation scheme, discussed in the main text, we employ to handle the higher-order fermionic interaction terms that arise from the long-range Heisenberg interactions. As explained in the main text, our approach is guided by the intuition that long-range couplings suppress local quantum fluctuations, allowing us to approximate nonlocal products of spin (or equivalently, fermionic) operators by their mean-field averages. This assumption greatly simplifies the resulting effective Hamiltonian and makes it tractable to derive self-consistent equations for the order parameters.

To characterize our mean-field state, we introduce two key parameters. First, we define a uniform magnetization and staggered magnetization along the $z$-axis, averaged over all lattice sites

\begin{align}
s^z &= \frac{1}{N}\bigg\langle \sum_{j} \left(\sigma_{j,A}^z + \sigma_{j,B}^z\right)\bigg\rangle, \label{eqsz} \\ n^z &= \frac{1}{N}\bigg\langle \sum_{j} \left(\sigma_{j,A}^z - \sigma_{j,B}^z\right)\bigg\rangle \label{eqnz} \ .
\end{align}

\noindent Here, $s^z$ captures the uniform component of the $z$-magnetization, while $n^z$ captures the alternating (Néel-type) order that is expected to develop in the presence of antiferromagnetic couplings and a staggered field.

To proceed, we consider states $|\psi\rangle$ that have the general properties we expect of the eigenstates of the Hamiltonian. In the main text we outlined the following assumptions 

\begin{enumerate}
    \item The state is characterized by $s^z$ \eqref{eqsz} and $n^z$ \eqref{eqnz}.

    \item The wavefunction fulfills bipartite permutational symmetry meaning $\braket{\sigma_{i,L}^z\sigma_{j, M}^z} = \braket{\sigma_{i', L}^z\sigma_{j',M}}$ where $i, i'$ and $j, j'$ are site indices and $L,M \in \{A, B\}$ are sublattice indices. Note this is reduced with respect to the full permutational symmetry of other long-range systems $\braket{\sigma_i^z \sigma_j^z} = \braket{\sigma_{i'}^z \sigma_{j'}^z}$~\cite{sugimoto_eigenstate_2022, defenu_metastability_2021}. 

    \item We assume mean field decoupling for the expectation values of the collective spins $\braket{S_A^z S_B^z} \approx \braket{S^z}^2 - \braket{N^z}^2$ . 

\end{enumerate}

\noindent As a concrete realization of such a state, we write down a variational wavefunction $|\psi\rangle$ built from a uniform superposition of four basis states arranged in pairs on the $A,B$ sublattices. Denoting spin-up ($|1\rangle$) and spin-down ($|0\rangle$) states, we consider

\begin{align}
|\psi\rangle = a|\mathbf{00}\rangle + b|\mathbf{01}\rangle + c|\mathbf{10}\rangle + d|\mathbf{11}\rangle,
\end{align}

\noindent where $|\mathbf{ij}\rangle$ represents a translationally invariant product state with every $A$-site spin in state $|i\rangle$ and every $B$-site spin in state $|j\rangle$. As an example $\ket{\boldsymbol{01}} = \ket{01 01 01 01 01 \dots}$. The coefficients $a, b, c, d$ are chosen to satisfy

\begin{enumerate}
\item  Normalization \(\langle \psi|\psi \rangle = 1\).
\item  Uniform $z$-magnetization \( s^z= \frac 1N \sum_j \braket{\psi|\sigma_{j,A}^z+\sigma_{j,B}^z|\psi} = 2(|d|^2 - |a|^2) \).
\item  Staggered $z$-magnetization \(n^z= \frac 1N \sum_j \braket{\psi|\sigma_{j,A}^z-\sigma_{j,B}^z|\psi} = |b|^2 - |c|^2 \).
\item  Correct two-point correlations \(|a|^2 - |b|^2 - |c|^2 + |d|^2 = (s^z)^2 - (n^z)^2\).
\end{enumerate}
Solving these conditions yields explicit expressions for $a, b, c, d$ in terms of $s^z$ and $n^z$. This construction ensures that $|\psi\rangle$ reproduces the desired magnetization and correlation pattern. \\

 \noindent Within our mean-field approximation, we replace string operators by their expectation values in the state $|\psi\rangle$. Since $|\psi\rangle$ was constructed to yield the correct uniform and staggered magnetization and their correlations, evaluating these string operators reduces to a straightforward computation.

For string operators of even length, we find that their expectation values factorize as follows

\begin{align}
\braket{T^{AB}_{i,j}} = \langle \psi|\prod_{m=i+1}^{i+r+1}\sigma_{m,B}^z \sigma_{m,A}^z|\psi \rangle = |a|^2 + |d|^2 + (-1)^{r}(|c|^2  + |d|^2)=
\begin{cases}
1 & \text{if $r$ is even\ ,}\\[6pt]
(s^z)^2-(n^z)^2 & \text{if $r$ is odd}\ .
\end{cases} 
\end{align}

\noindent where we introduced $r = i-j$. Alternatively, one can write $\braket{T_{i,j}^{AB}}  = (1 + (-1)^r)/2  + (1 - (-1)^r)/2 ((s^z)^2 - (n^z)^2)$. Next we can evaluate the expectation value  

\begin{align}
\braket{T^{BA}_{i,j}} = \langle \psi|\prod_{m=i+1}^{i+r+1}\sigma_{m,A}^z \sigma_{m,B}^z|\psi \rangle = |a|^2 + |d|^2 + (-1)^{r}(|c|^2  + |b|^2)=
\begin{cases}
1 & \text{if $r$ is even,}\\[6pt]
(s^z)^2-(n^z)^2 & \text{if $r$ is odd}\ .
\end{cases} 
\end{align}

\noindent Next, we can evaluate the string operators for interaction in one sublattice. Specifically we find 

\begin{align}
\braket{T^{AA}_{ij}} &= \bigg\langle\sigma_{i+1,B}^z\bigg(\prod_{m=i+2}^{j-1}\sigma_{m,B}^z \sigma_{m,B}^z\bigg) \bigg\rangle = |a|^2 +|d|^2 - (-1)^r(|c|^2 - |b|^2) = s^z-(-1)^r n^z \ ,
\\
    \braket{T_{ij}^{BB}} &= \bigg\langle\bigg(\prod_{m=i+1}^{j-2} \sigma_{m,A}^z \sigma_{m,B}^z\bigg) \sigma_{j-1,A}^z\bigg\rangle = |a|^2 +|d|^2 + (-1)^r(|c|^2 - |b|^2) = s^z + (-1)^r n^z \ .
\end{align}

\noindent Using these expressions, we can write the $xy$ part of the inteaction in the following way 

\begin{align}
    H_{xy} = \frac J {4Z_\alpha} \sum_{i =1}^N \sum_{r=1}^{N/2} \frac 1 {r_{ij}} \bigg( 2\braket{T_{i,i+r}^{AB}}(c_{i, A}^\dagger c_{i+r, B } + \mathrm{h.c.}) + \braket{T^{AA}_{i,i+r}}(c_{i,A}^\dagger c_{i+r,A} + \mathrm{h.c.} ) + \braket{T^{BB}_{i,i+r}}(c_{i,B}^\dagger c_{i+r,B} + \mathrm{h.c.} )  \bigg)  \ .
\end{align}

Nevertheless, the mean-field expressions for the string operators are still quite complicated. However, we can simplify them when $s^z\approx 0$ and $n^z \approx -1$. We define $\Delta = (n^z)^2 - (s^z)^2$ and note $\Delta \approx 1$. In that case we can simplify

\begin{align}
    \braket{T^{AB}_{ij}}   = \frac 1 2 (1 - \Delta ) + \frac{(-1)^r}{2}(1+\Delta) \approx \frac{(-1)^r} 2 (1 + \Delta) \ ,
\end{align}

\noindent  Furthermore, we can approximate $\braket{T_{ij}^{AA}} \approx -(-1)^r n^z$ and $\braket{T_{ij}^{BB}} \approx (-1)^r n^z$ The Hamiltonian simplifies to 

\begin{align}
    H_{xy} =\frac J {4Z_\alpha} \sum_{i =1}^N \sum_{r=1}^{N/2} \frac {(-1)^r} {r_{ij}^\alpha} (1+\Delta)(c_{i, A}^\dagger c_{i+r, B } + c_{i,B}^\dagger c_{i+r, A} + \mathrm{h.c.}) - 2n^z(c_{i,B}^\dagger c_{i+r,B} -c_{i,A}^\dagger c_{i+r,A} + \mathrm{h.c.} )   \bigg)   \ .\label{eq:HamiltonianIntermed}
\end{align}

\noindent Next, we focus on the $zz$ interaction between the spins. Recall that the original $zz$-interaction term is given by
\begin{align}
    H_{zz} &= \frac J {4Z_\alpha} \sum_i \sum_r \frac{1}{r_{ij}^\alpha} (\sigma_{i,A}^z + \sigma_{i,B}^z)(\sigma_{i+r,A}^z + \sigma_{i+r,B}^z) \,.  \label{eq:Hzzbare}
\end{align}

\noindent To proceed with a mean-field approximation, we assume that fluctuations around the average values of these spin-like operators are small. We introduce the following average quantities
\begin{align}
    \langle \sigma_{j,A}^z \rangle = s^z + n^z \,, \qquad
    \langle \sigma_{j,B}^z \rangle = s^z - n^z \,.
\end{align}
\noindent Here, $s^z$  \eqref{eqsz}  represents the average symmetric part (the uniform $z$-component of spin), and $n^z$ \eqref{eqnz} represents the average antisymmetric part (the sublattice imbalance in the $z$-component of spin). These parameters $s^z$ and $n^z$ are determined self-consistently. With these definitions,
\begin{align}
    \langle \sigma_{j,A}^z + \sigma_{j,B}^z \rangle 
    &= \langle \sigma_{j,A}^z \rangle + \langle \sigma_{j,B}^z \rangle 
    = (s^z + n^z) + (s^z - n^z) = 2 s^z \,.
\end{align}

\noindent This means the simplest possible approximation is 

\begin{align}
    H_{zz} &= JN(s^z)^2  \,.  
\end{align}

\subsection{\texorpdfstring{Fourier Transform of the Hamiltonian}{Fourier Transform of the Hamiltonian}}

\noindent We now introduce discrete Fourier transforms for the fermionic operators
\begin{align}
c_{j,A/B} &= \frac{1}{\sqrt{N}}\sum_{n=0}^{N-1} e^{- i \frac{2\pi}{N} n j} c_{n,A/B}, \qquad
c_{n,A/B} = \frac{1}{\sqrt{N}}\sum_{j=0}^{N-1} e^{ i \frac{2\pi}{N} n j} c_{j,A/B}, 
\end{align}
with $n \in \{0,\dots,N-1\}$. 
\noindent We will first apply this transform to the $xy$ coupling term between the sublattices \eqref{eq:HamiltonianIntermed}
\begin{align}
H_{xy}^{\mathrm{inter}} &= \frac J {4Z_\alpha} \sum_{i=0}^{N-1} \sum_{r=1}^{N/2} \frac{(-1)^r}{r^\alpha}
(1+\Delta)\bigl(c_{i,A}^\dagger c_{i+r,B} + c_{i,B}^\dagger c_{i+r,A} + \mathrm{h.c.}\bigr) \\
&= \frac J 2 (1+\Delta) \sum_{n=0}^{N-1} F_\alpha(n)\left[ \  c_{n,A}^\dagger c_{n,B}
+ \ c_{n,B}^\dagger c_{n,A} \right] ,
\end{align}
where we have defined the form factor
\begin{align}
F_{\alpha}(n) \equiv \frac 1 {Z_\alpha} \sum_{r=1}^{N/2} \frac{(-1)^r}{r^\alpha} \cos\bigg( 2\pi   \frac rN n\bigg) \ .
\label{eq:Falpha_n}
\end{align}

\noindent Next, we turn our attention to the intra sublattice coupling terms in Eq. \eqref{eq:HamiltonianIntermed}
\begin{align}
H_{xy}^{\mathrm{intra}} &=  \frac {Jn^z} {2Z_\alpha}  \sum_{i=0}^{N-1} \sum_{r=1}^{N/2}\frac{(-1)^r}{r^\alpha}\bigl(c_{i,A}^\dagger c_{i+r,A}-c_{i,B}^\dagger c_{i+r,B}+\mathrm{h.c.}\bigr) \\ &= -J n^z \sum_{n=0}^{N-1} F_\alpha(n) ( c_{n, A}^\dagger c_{n, A} - c_{n, B}^\dagger c_{n,B}).
\end{align}

\noindent where $\epsilon_\alpha(n) = \mathrm{Re}[F_\alpha(n)]$. Finally, the Zeeman term transforms straightforwardly
\begin{align}
\begin{split}
H_{\mathrm{mag}} &= -2h\sum_{i=0}^{N-1}(c_{i,A}^\dagger c_{i,A} - c_{i,B}^\dagger c_{i,B})
= -2h \sum_{n=0}^{N-1} (c_{n,A}^\dagger c_{n,A} - c_{n,B}^\dagger c_{n,B}).
\end{split}
\end{align}

\noindent In summary, writing the entire Hamiltonian we arrive at Eq.~(4) of the main text 
\begin{align}
\boxed{
H  = JN(s^z)^2 - 2\sum_{n=0}^{N-1} \begin{pmatrix} c_{A, n}^\dagger \\ c_{B, n}^\dagger \end{pmatrix}
\begin{pmatrix}
 h_{N, n}(n^z) & \Gamma_n(s^z, n^z) \\
\Gamma_n^*(s^z, n^z) & - h_{N, n}(n^z)
\end{pmatrix}
\begin{pmatrix} c_{A, n} \\ c_{B, n} \end{pmatrix}\ ,}\  \label{eq:mean_field_Hamiltonian_n}
\end{align}
\noindent where the $2 \times 2$ Hamiltonian in momentum space self-consistently depends on the collective observables $s^z$ and $n^z$  as follows
\begin{align}
h_{N, n}(n^z) &= h_N + J n^z \epsilon_\alpha (n), \label{eq:hN_n} \\
\Gamma_n(s^z, n^z) &= - \frac J2 \left(1 + (n^z)^2 - (s^z)^2 \right) F_\alpha(n)\ . \label{eq:Gamma_n}
\end{align}

\noindent It is important to note that the approximation $(n^z)^2 \approx 1$ is employed to ensure the validity of the mean-field Hamiltonian. Deviations from this approximation would invalidate the derivation of the Hamiltonian.

To diagonalize the Hamiltonian, we introduce new fermionic operators $d_{\pm, n}$ and $d_{\pm, n}^\dagger$, which satisfy the canonical anticommutation relations
\begin{equation}
\{d_{s, n'}, d_{s', n}^\dagger\} = \delta_{n n'} \delta_{s s'}. \label{eq:anticommutation_relations}
\end{equation}
We perform a unitary transformation to these new operators via the following ansatz
\begin{equation}
\begin{pmatrix} d_{n} \\ d_{-n} \end{pmatrix} 
= \begin{pmatrix} 
\cos\left(\frac{\theta_n}{2}\right) & -\sin\left(\frac{\theta_n}{2}\right) \\ 
\sin\left(\frac{\theta_n}{2}\right) & \cos\left(\frac{\theta_n}{2}\right) 
\end{pmatrix} 
\begin{pmatrix} c_{n, A} \\ c_{n, B} \end{pmatrix}. \label{eq:fermionic_transformation}
\end{equation}
Applying this transformation to the Hamiltonian and requiring that off-diagonal terms vanish leads to the condition
\begin{equation}
\tan(\theta_n) = \frac{\Gamma_n}{h_{N, n}}. \label{eq:theta_condition}
\end{equation}
Solving the above condition, the eigenenergies of the system are obtained as
\begin{equation}
E_{A/B, n} =   \pm \sqrt{h_{N, n}^2 + |\Gamma_n|^2} . \label{eq:eigenenergies}
\end{equation}

\noindent We note the spectrum of the Eigenenergies must be evaluated carefully. Specifically it depends on the form factor \eqref{eq:Falpha} 

\begin{align}
    F_{\alpha}(n) = (1-\alpha) 2^{1-\alpha  } \sum_{r=1}^{N/2}\frac 1 N  \frac{(-1)^r}{(r/N)^\alpha} \cos\bigg( 2\pi   \frac rN n\bigg)  \ . \label{eq:Falpha}
\end{align}

\noindent In the thermodynamic limit, the normalization factor \eqref{eq:normalization} diverges, meaning that the limit in equation \eqref{eq:Falpha} must be taken carefully. We can express the equation in terms of the continuous variable $s = r/N$, giving us Eq.~(5) of the main text 

\begin{align}
    \boxed{ F_\alpha(n)  = (1-\alpha)2^{1- \alpha}\int_0^{1/2}ds\ \frac{\cos(2\pi s n)}{s^\alpha} \ . }
\end{align} 

\noindent where we used the Riemann integral formula and shifted the indices by $n\to n + N/2$ to absorb the oscillating $(-1)^r$ factor.
 This shows in the case $\alpha < 1$ the spectrum \eqref{eq:eigenenergies} of the Hamiltonian becomes discrete. This is the case as the energy difference $\epsilon_{\alpha}(n+1) - \epsilon_\alpha(n)$ stays finite as $N\to \infty$.

In the transformed basis \eqref{eq:fermionic_transformation}, the collective operators can be expressed as
\begin{align}
s^z &= \frac 1 N \sum_j \sigma_{j, A}^z + \sigma_{j, B}^z = 1 - \frac 2N \sum_{n\neq 0}  d_n^\dagger d_n, \label{eq:S_z} \\
n^z &= \frac 2N \sum_{n > 0} \cos(\theta_n) \left(d_n^\dagger d_n - d_{-n}^\dagger d_{-n}\right) - \sin(\theta_n) \left(d_n^\dagger d_{-n} + d_{-n}^\dagger d_n\right). \label{eq:N_z}
\end{align}

Through this mean-field approach, we have derived a self-consistent Hamiltonian that incorporates the collective observables $s^z$ and $n^z$. By performing a unitary transformation to diagonalize the Hamiltonian, we obtain simple expressions for the eigenenergies and the collective operators in the new fermionic basis. This framework facilitates further analysis of the system's properties and the exploration of its quantum phases.

\subsection{\texorpdfstring{Spectrum in the $\alpha = 0$ case}{Spectrum in the alpha = 0 case}}

In the main text in Eq. (4), we have introduced a form factor $F_\alpha(n)$, which determines the structure of the long-range Hamiltonian. For $\alpha<1$, the energy differences $\epsilon_{\alpha}(n+1)-\epsilon_{\alpha}(n)$ remain finite, and the single-particle spectrum is discrete even in the thermodynamic limit. When $\alpha=0$, the behavior of $F_0(n)$ simplifies dramatically, and in the low-energy regime, we make the approximation that $n^z \approx -1$. Under this assumption, the form factor $F_0(n)$ takes on a particularly simple form
\begin{align}
F_0(n) = \delta_{|n|,1}.
\end{align}
This approximation implies that the parameters \eqref{eq:hN_n} and \eqref{eq:Gamma_n} in the Hamiltonian \eqref{eq:mean_field_Hamiltonian_n} become 
\begin{align}
h_{N, n}(n^z) &= h + J n^z \delta_{|n|,1},  \\
\Gamma_n(s^z, n^z) &= - \frac J2 \left(1 + (n^z)^2 - (s^z)^2 \right) \delta_{|n|,1}\ . 
\end{align}

\noindent Therefore, we obtain the following single-particle eigenenergies using $E_{\pm n} = \pm \sqrt{h_{n,N}^2 + \Gamma_n^2 } $

\begin{align}
E_{n} =\begin{cases}   \pm \lambda_1 \sqrt{(h+J n^z)^2 + (\frac{J}2)^2 (1+(n^z)^2 -(s^z)^2 )^2 } \qquad &|n|= 1 \ ,\\ \pm\lambda_n    h &|n|>1 \ . \end{cases}  \label{eq:Ealpha0_supp}
\end{align}

\noindent where we have introduced the rescaling factors $\lambda_n$. These factors will ensure that the energy is extensive. We now construct the many-body spectrum. For this, the single-particle energy levels are occupied with fermions. The energy of the many-body state is given by

\begin{align}
    \braket{H} = \sum_{n\neq 0} E_n \braket{d_n^\dagger d_n} \ , \label{eq:manyBodyEnergy}
\end{align}

\noindent with $E_n$ according to equation \eqref{eq:Ealpha0_supp}. $E_n$ depends on the collective observables $s^z$ and $n^z$. $s^z$ is a conserved quantity and is uniquely determined by the magnetization sub-sector according to Eq. \eqref{eq:Sz}.

\noindent However, $n^z$ is not a conserved quantity and requires a self-consistent determination via

\begin{align}
    n^z &= \frac 2N \sum_{n > 0} \cos(\theta_n(s^z, n^z)) \braket{d_n^\dagger d_n - d_{-n}^\dagger d_{-n}} \ , \label{eq:selfconsistent0}
\end{align}

\noindent where the Bogoliubov angles are given by

\begin{align}
\cos(\theta_n) = \frac{h_{N,n}}{\sqrt{h_{N,n}^2 + \Gamma_n^2}} = \begin{cases} \frac{h + Jn^z}{\sqrt{(h + Jn^z)^2 + (\frac{J}2)^2(1+(n^z)^2 - (s^z)^2 )^2 }} \qquad &|n| = 1  \\ 1 &|n|>1\end{cases} \label{eq:bogo_supp} \ \ .
\end{align}

To determine $n^z$ self-consistently for given occupation numbers $\{\braket{d_n^\dagger d_n}\}$ characterizing an excited state, we employ an iterative approach. Starting with an initial guess for $n^z$ (e.g., $n^z = 0$), we update it iteratively. In each iteration, the current value of $n^z$ and the fixed value of $s^z$ for a given magnetization sector are used to compute the single-particle energies $E_n$ according to equation \eqref{eq:Ealpha0_supp}. The occupation numbers $\braket{d_n^\dagger d_n}$ are given and fixed. We then update the value of $n^z$ using equation \eqref{eq:selfconsistent0} and the given occupation numbers. This iterative process is continued until the value of $n^z$ converges, meaning that the difference between successive iterations falls below a predefined tolerance. The converged $n^z$ represents the self-consistent solution for the given occupation numbers and magnetization sector.

Alternatively, we can approximate $n^z$ for large large $N$. Substituting equation \eqref{eq:bogo_supp} into \eqref{eq:selfconsistent0}, we get:

\begin{align}
n^z = \frac 2N \left( \frac{h + Jn^z}{\sqrt{(h + Jn^z)^2 + \frac{J^2}{4}(1+(n^z)^2 - (s^z)^2 )^2 }} \braket{d_1^\dagger d_1 - d_{-1}^\dagger d_{-1}} + \sum_{n>1} \braket{d_n^\dagger d_n - d_{-n}^\dagger d_{-n}} \right) \ .
\end{align}

\noindent In the thermodynamic limit ($N \rightarrow \infty$), the prefactor $2/N$ suppresses the first term, as it only contains the $n=\pm 1$ contributions. Thus, we can approximate the self-consistency

\begin{align}
n^z \approx \frac{2}{N} \sum_{n>1}  \braket{d_n^\dagger d_n - d_{-n}^\dagger d_{-n}} \ . \label{eq:selfconsistent0_approx}
\end{align}

\noindent This approximation is valid for large systems and when the external field $h$ significantly exceeds the interaction strength $J$. This is the approxmiation we use in the main text. 

Given the factors $\lambda_n$, the energy must also be determined self-consistently by demanding the many-body energies $E(N, s^z,n^z)$ are extensive $E(N, s^z, n^z) = NE(1, s^z, n^z)$.  Using the self-consistency equation \eqref{eq:selfconsistent0_approx} and substituting \eqref{eq:Ealpha0_supp} into equation \eqref{eq:manyBodyEnergy} we find 

\begin{align}
    E &= JN(s^z)^2  - h\sum_{n>2} \lambda_n\braket{d_n^\dagger d_n -d_{-n}^\dagger d_{-n}} -\lambda_1 \sqrt{(h+ Jn^z)^2 + \frac{J^2}4 (1+ (n^z)^2 - (s^z)^2)^2}
\end{align}

\noindent We recognize $\sum_n \braket{d_n^\dagger d_n + d_{-n}^\dagger d_{-n}} = Ns^z$ and $\sum_{n} \braket{d_n^\dagger d_n - d_{-n}^\dagger d_n} \approx N n^z$. Furthermore, from demanding extensivity $E(\mu N, s^z, n^z) = \mu E( N, s^z, n^z)$ from the many-body energies we find $\lambda_n  = 1$ for $n>1$ and $\lambda_1 = N$. Using this we find the following energies (neglecting constant shifts)

\begin{align}
    E(n^z, s^z)  =  h N n^z  + J N(s^z)^2  -  N\sqrt{(h + J n^z)^2 + \frac{J^2}4\left(1+ (n^z)^2 - (s^z)^2\right)^2} \ , \label{eq:E_of_nz_sz}
\end{align}

\noindent where we have expressed the energy in terms of $s^z = S^z/N$ and $n^z=N^z/N$.
This expression represents the total energy of the system as a function of the conserved quantity $s^z$ and the self-consistently determined quantity $n^z$.

We highlight that different energy levels are generally \emph{incommensurable}. The dimensionless quantities \(n^z\) and \(s^z\) vary in increments of \(1/N\), and substituting them into Eq.~\eqref{eq:E_of_nz_sz} reveals that each energy level is derived from the square root of a sum of squares involving polynomials in \(n^z\) and \(s^z\). Specifically,

\begin{align}
    E(n^z, s^z)  =  h n_1   + J n_2^2/N  -  N\sqrt{(h + J n_1/N)^2 + \frac{J^2}4\left(1+ \left(\frac{n_1}N\right)^2 - \left(\frac{n_2}N\right)^2\right)^2} \ .
\end{align}

For generic real values of the system parameters \(h\) and \(J\), the expressions inside the square root do not simplify to ratios of integers or other rational numbers. The presence of the square root of a sum of squares typically introduces algebraic irrationals, preventing the energy levels from forming simple rational ratios. Additionally, the coupling between \(n^z\) and \(s^z\) through the term \(\left(1 + (n^z)^2 - (s^z)^2\right)^2\) further complicates the relationship between different energy levels, making systematic degeneracies or commensurate patterns highly unlikely without specific fine-tuning.

In cases where the ratio \(h/J\) is rational, such as \(h/J = p/q\) with integers \(p\) and \(q\), there may exist subsets of energy levels that exhibit partial commensurability or isolated degeneracies. Similarly, in the limit where \(h = 0\) or \(h\) is extremely large compared to \(J\), the expression for \(E_{n_1,n_2}\) simplifies but does not generally lead to a commensurate spectrum across all \((n_1, n_2)\) pairs. Specific integer values of \(n_1\) and \(n_2\) might produce individual energy levels with simpler forms, but these instances are exceptional and do not affect the overall incommensurability of the spectrum.

Therefore, except for these isolated or fine-tuned scenarios, the set of energy levels \(\{E_{n_1,n_2}\}\) does not admit a representation in terms of simple rational multiples. The intricate dependence on both \(n^z\) and \(s^z\) within the square root ensures that the energy levels are incommensurate for generic \(h\) and \(J\). This incommensurability implies a quasicontinuous spectrum without simple rational structures.

\section{Key expectation values in the initial states}

In the main text, we discuss the time evolution for both the coherent Néel state as well as the mixed Néel state. In this section, we will explicitly derive the expectation values of key observables in these two initial states. 

\subsection{Observables in coherent Néel states}

The goal of this section is to compute the expectation values of key observables in the coherent Néel states forming the mixed Néel state ensemble. According to the main text the coherent Néel state is defined as

\begin{align}
\label{eq:coherent_A}
\ket{\vartheta}_{j,A} &= \cos(\vartheta/2) \ket{1}_{j,A}  + \sin(\vartheta/2)\ket{0}_{j, A}\ , \\
\label{eq:coherent_B}
\ket{\vartheta}_{j,B} &= \sin(\vartheta/2) \ket{1}_{j,B}  - \cos(\vartheta/2) \ket{0}_{j, B} \ , 
\end{align}

\noindent where $\ket{0}$ and $\ket{1}$ are the eigenstates of the $\sigma^z$ operator, representing spin down and spin up, respectively. The total coherent state is a product state over all pairs

\begin{equation}
\label{eq:total_coherent}
\ket{\vartheta} = \bigotimes_j \ket{\vartheta}_{j,A} \otimes \ket{\vartheta}_{j,B} \ .
\end{equation}

\noindent The goal of this section is to compute the expectation values of the following observables: the total spin $z$-component, defined as \( S^z = \sum_j \sigma_{j,A}^z + \sigma_{j,B}^z \); the spin imbalance $z$-component, given by \( N^z = \sum_j \sigma_{j,A}^z - \sigma_{j,B}^z \); the square of the total spin $z$-component, \((S^z)^2\); and the square of the spin imbalance $z$-component, \((N^z)^2\).

First, we calculate the expectation value of $\sigma_{j,A}^z$ and $\sigma_{j,B}^z$ for a single pair using \eqref{eq:coherent_A} and \eqref{eq:coherent_B}

\begin{align}
\label{eq:sigma_zA}
\langle \vartheta | \sigma_{j,A}^z | \vartheta \rangle_{j,A} &= \cos^2(\vartheta/2) - \sin^2(\vartheta/2) = \cos(\vartheta) \ , \\
\label{eq:sigma_zB}
\langle \vartheta | \sigma_{j,B}^z | \vartheta \rangle_{j,B} &= \sin^2(\vartheta/2) - \cos^2(\vartheta/2) = -\cos(\vartheta) \ .
\end{align}

\noindent Since the total state is a product state \eqref{eq:total_coherent}, the expectation value of $S^z$ is the sum of the expectation values for each pair. This means we find for the collective observables

\begin{align}
\langle S^z \rangle 
&= \sum_j (\langle \vartheta | \sigma_{j,A}^z | \vartheta \rangle_A + \langle \vartheta | \sigma_{j,B}^z | \vartheta \rangle_B ) 
= 0 \label{eq:Sz} \ , \\ 
\langle N^z \rangle 
&= \sum_j (\langle \vartheta | \sigma_{j,A}^z | \vartheta \rangle_A - \langle \vartheta | \sigma_{j,B}^z | \vartheta \rangle_B) 
= N \cos(\vartheta) \ .\label{eq:nz}
\end{align}

\noindent The expectation value of the spin imbalance is proportional to the number of spin pairs and $\cos(\vartheta)$. We now compute the expectation value of $(S^z)^2$

\begin{equation}
(S^z)^2 = \left( \sum_j \sigma_{j,A}^z + \sigma_{j,B}^z \right) \left( \sum_k \sigma_{k,A}^z + \sigma_{k,B}^z \right) = \sum_{j,k} \left( \sigma_{j,A}^z \sigma_{k,A}^z + \sigma_{j,A}^z \sigma_{k,B}^z + \sigma_{j,B}^z \sigma_{k,A}^z + \sigma_{j,B}^z \sigma_{k,B}^z \right) \ .
\end{equation}

\noindent First we consider the diagonal terms ($j=k$).
Using the fact that $(\sigma_{j,\alpha}^z)^2 = I$ (the identity operator), we have $\langle \vartheta | (\sigma_{j,A}^z)^2 | \vartheta \rangle = 1 $ and  $\langle \vartheta | (\sigma_{j,B}^z)^2 | \vartheta \rangle = 1$
Using \eqref{eq:sigma_zA} and \eqref{eq:sigma_zB}
\begin{align}
    \langle \vartheta | \sigma_{j,A}^z \sigma_{j,B}^z | \vartheta \rangle = -\cos^2(\vartheta) \ .
\end{align}
\noindent Thus the diagonal contribution for each $j$ is $2(1-\cos^2\vartheta)$. Summing over all $N$ pairs gives a total diagonal contribution of $2N(1 - \cos^2(\vartheta)) = 2N \sin^2(\vartheta)$.

Next we consider the off-diagonal contribution $j \neq k$. 
Since the state is a product state, we can factorize the expectation values and find $\langle \vartheta | \sigma_{j,A}^z \sigma_{k,A}^z | \vartheta \rangle = \langle \vartheta | \sigma_{j,B}^z \sigma_{k,B}^z | \vartheta \rangle = \cos^2(\vartheta)$ for the correlator inside one sublattice and  
$\langle \vartheta | \sigma_{j,A}^z \sigma_{k,B}^z | \vartheta \rangle = -\cos^2(\vartheta)$ between the sublattices. The total contribution for each pair $j \neq k$ is therefore $\cos^2 \vartheta - \cos^2 \vartheta - \cos^2 \vartheta + \cos^2 \vartheta = 0$. Combining the diagonal and off-diagonal terms
\begin{equation}
    \langle (S^z)^2 \rangle = 2N \sin^2(\vartheta) \ .
\end{equation}

\noindent The expectation value of $(S^z)^2$ scales linearly with $N$ due to the cancellation of off-diagonal terms.

Finally, we calculate the expectation value of $(N^z)^2$

\begin{equation}
(N^z)^2 = \left( \sum_j \sigma_{j,A}^z - \sigma_{j,B}^z \right) \left( \sum_k \sigma_{k,A}^z - \sigma_{k,B}^z \right) = \sum_{j,k} \left( \sigma_{j,A}^z \sigma_{k,A}^z - \sigma_{j,A}^z \sigma_{k,B}^z - \sigma_{j,B}^z \sigma_{k,A}^z + \sigma_{j,B}^z \sigma_{k,B}^z \right) \ .  \label{eq:NzMult}
\end{equation}

Equation \eqref{eq:NzMult} contains diagonal terms with $j= k$ and off-diagonal terms with $j \neq k$. We first calculate the expectation value for the diagonal terms 
($j=k$)
\begin{align}
\langle \vartheta | (\sigma_{j,A}^z)^2 + (\sigma_{j,B}^z)^2 - 2\sigma_{j,A}^z\sigma_{j,B}^z | \vartheta \rangle = \begin{cases}  2(1+\cos^2 \vartheta)  \qquad &\mathrm{for} \qquad j = k \\ 4\cos^2(\vartheta) \qquad &\mathrm{for} \qquad j\neq k \ .\end{cases} 
\end{align}
\noindent Keeping in mind that there are in total $N$ diagonal terms and $N(N-1)$ off-diagonal terms we find 
\begin{equation}
    \label{eq:Nz2}
\begin{aligned}
\langle (N^z)^2 \rangle = 4N^2 \cos^2(\vartheta) + 2N \sin^2(\vartheta) \ .
\end{aligned}
\end{equation}

\noindent The expectation value of $(N^z)^2$ has both a quadratic and a linear dependence on $N$. Neverthless, the variance in $N^z$ only depends linearly on $N$

\begin{align}
    (\Delta N^z)^2 = \braket{(N^z)^2} - \braket{N^z}^2 = 2N \sin^2(\vartheta) \ . 
\end{align}

\subsection{Observables in mixed Néel states}

This section details the calculation of the expectation value of $N^z$ in the mixed Néel state using a basis rotation to simplify the integral. The mixed Néeel state coherent state is described by the density matrix (as defined in Equation (2) of the main text)

\begin{equation}
\label{eq:density_matrix_supp}
\rho(\Omega_0, \sigma) = \frac{1}{Z} \int d\Omega \, e^{-\frac{(1 - \mathbf{r}(\Omega) \cdot \mathbf{r}(\Omega_0))^2}{2\sigma^2}} \ket{\vartheta, \varphi}\bra{\vartheta, \varphi} \ , 
\end{equation}

\noindent where the normalization constant $Z$ is given by $Z = \sqrt{2\pi^3} \mathrm{erf}(\sqrt{2}/\sigma) \sigma$. The expectation value of $N^z$ is given by

\begin{equation}
\label{eq:Nz_rho_integral}
\langle N^z \rangle_\rho = \frac{1}{Z} \int d\Omega \, e^{-\frac{(1 - \mathbf{r}(\Omega) \cdot \mathbf{r}(\Omega_0))^2}{2\sigma^2}} (2N \cos(\vartheta)) \ .
\end{equation}

\noindent Here, $\Omega = (\vartheta, \varphi)$ represents the spherical coordinates, and $\mathbf{r}(\vartheta, \varphi)$ is the corresponding unit vector on the Bloch sphere

\begin{equation}
\label{eq:r_vector}
\mathbf{r}(\vartheta, \varphi) = (\sin(\vartheta) \cos(\varphi), \sin(\vartheta)\sin(\varphi), \cos(\vartheta)) \ . 
\end{equation}

We consider the case where $\Omega_0 = (\vartheta_0, 0)$, which means the reference direction $\mathbf{r}(\Omega_0)$ lies in the $xz$-plane. To simplify the integral in Equation \eqref{eq:Nz_rho_integral}, we perform a basis rotation such that the new $z$-axis aligns with $\mathbf{r}(\Omega_0)$. In this rotated frame, $\mathbf{r}(\Omega_0) = \hat{\mathbf{e}}_z = (0, 0, 1)$, and the dot product simplifies to

\begin{equation}
\label{eq:simplified_dot_product}
\mathbf{r}(\Omega) \cdot \mathbf{r}(\Omega_0) = \cos(\vartheta') \ , 
\end{equation}

\noindent where $\vartheta'$ is the polar angle in the rotated coordinate system.

We also need to transform the $2N\cos(\vartheta)$ term in the integrand. In the rotated frame we have that $\mathbf{r}(\Omega_0) = (0,0,1)$, this means we have $\vartheta_0 = 0$ for this rotated frame. From Equation (14) in the main text, 

\begin{align}
    \braket{N^z} \to \braket{N^z \cos(\vartheta_0) + N^y \sin(\vartheta_0)} = N \cos(\vartheta) \cos(\vartheta_0) + N \sin(\vartheta) \sin(\varphi) \sin(\vartheta_0) \ .
\end{align}

\noindent Substituting Equation \eqref{eq:simplified_dot_product} into Equation \eqref{eq:Nz_rho_integral} and performing the integration, we obtain the following result

\begin{equation}
\label{eq:Nz_rho_result}
\langle N^z \rangle_\rho = 2N \cos(\vartheta_0) \bigg( 1 - \sqrt{\frac{2}{\pi}}\frac{1 - e^{-2/\sigma^2}}{\mathrm{erf}(\sqrt 2/\sigma)}\bigg) \ .
\end{equation}

\noindent In the limit of $\sigma \to 0$ this equation simplifies to $\lim_{\sigma \to 0} \langle N^z \rangle_\rho = 2N \cos(\vartheta_0)$. This corresponds to the case of a pure coherent state aligned with $\Omega_0$, as expected.
On the other hand, for $\sigma \to \infty$ (maximal uncertainty) we find $\lim_{\sigma \to \infty} \langle N^z \rangle_\rho = 0$. In this limit, the uncertainty is maximal, and the expectation value of $N^z$ vanishes.

The expectation value of $(N^z)^2$ is given by

\begin{equation}
\label{eq:Nz2_rho_integral}
\langle (N^z)^2 \rangle_\rho = \frac{1}{Z} \int d\Omega \, e^{-\frac{(1 - \mathbf{r}(\Omega) \cdot \mathbf{r}(\Omega_0))^2}{2\sigma^2}} \langle \vartheta, \varphi | (N^z)^2 | \vartheta, \varphi \rangle \ .
\end{equation}

\noindent Here, $\Omega = (\vartheta, \varphi)$ represents the spherical coordinates, and $\mathbf{r}(\Omega)$ is the corresponding unit vector on the Bloch sphere. We again apply the basis rotation to simplify the integral. $(N^z)^2$ transforms as 

\begin{equation}
\label{eq:Nz2_transform}
(N^z)^2 \rightarrow (N^z \cos(\vartheta_0) + N^y \sin(\vartheta_0))^2 = (N^z)^2 \cos^2(\vartheta_0) + 2 N^z N^y \cos(\vartheta_0)\sin(\vartheta_0) + (N^y)^2 \sin^2(\vartheta_0) \ .
\end{equation}

We need to calculate the expectation values of $(N^z)^2$, $(N^y)^2$, and $N^z N^y$ in the rotated frame, where the state is a coherent state aligned with the new $z$-axis, denoted by $\ket{\vartheta', \varphi'} = \ket{0,0}$. We use the notation $\langle \dots \rangle'$ to denote expectation values in this rotated frame. We have
\begin{align}
    (N^z)^2 &= \left(\sum_j \sigma^z_{j,A} - \sigma^z_{j,B}\right)^2 = \sum_{j,k} \left(\sigma^z_{j,A} \sigma^z_{k,A} - \sigma^z_{j,A} \sigma^z_{k,B} - \sigma^z_{j,B} \sigma^z_{k,A} + \sigma^z_{j,B} \sigma^z_{k,B} \right) \label{eq:Nz2_expanded} \ , \\
    (N^y)^2 &= \left(\sum_j \sigma^y_{j,A} - \sigma^y_{j,B}\right)^2 = \sum_{j,k} \left(\sigma^y_{j,A} \sigma^y_{k,A} - \sigma^y_{j,A} \sigma^y_{k,B} - \sigma^y_{j,B} \sigma^y_{k,A} + \sigma^y_{j,B} \sigma^y_{k,B} \right) \label{eq:Ny2_expanded}\ , \\
    N^z N^y &= \left( \sum_j \sigma^z_{j,A} - \sigma^z_{j,B} \right)\left( \sum_k \sigma^y_{k,A} - \sigma^y_{k,B} \right) = \sum_{j,k} \left( \sigma^z_{j,A} \sigma^y_{k,A} - \sigma^z_{j,A}\sigma^y_{k,B} - \sigma^z_{j,B}\sigma^y_{k,A} + \sigma^z_{j,B} \sigma^y_{k,B} \right) \ . \label{eq:NzNy_expanded}
\end{align}

\noindent For evaluating these correlators, we have $\langle \sigma_{j, A/B}^x \rangle = \pm \sin(\vartheta) \cos(\varphi)$, $\langle \sigma_{j, A/B}^y \rangle = \pm \sin(\vartheta) \sin(\varphi)$, and $\langle \sigma_{j, A/B}^z \rangle = \cos(\vartheta)$. We find

\begin{align}
    \langle (N^z)^2 \rangle' &= 4N^2 \cos^2(\vartheta) + 2N \sin^2(\vartheta) \ .
\end{align}

We can next focus on the correlator $\braket{(N^y)^2}$. We find for the diagonal and off-diagonal entries

\begin{align}
    \braket{\sigma^y_{j,A} \sigma^y_{k,A} - \sigma^y_{j,A} \sigma^y_{k,B} - \sigma^y_{j,B} \sigma^y_{k,A} + \sigma^y_{j,B} \sigma^y_{k,B} } =\begin{cases} 2 + 2\sin^2(\vartheta)\sin^2(\varphi)  \qquad &\mathrm{for} \qquad j=k \\ 4\sin^2(\vartheta)\sin^2(\varphi)  \qquad &\mathrm{for} \qquad j \neq k  \ . \end{cases} 
\end{align}

\noindent Combining these results we find 

\begin{align}
    \braket{(N^y)^2 } = 2N(1 - \sin^2(\vartheta)\sin^2(\varphi) ) + 4N^2 \sin^2(\vartheta) \sin^2(\varphi) \ .
\end{align}

\noindent In the case $\varphi = \pi/2$ this expression simplifies to $\braket{(N^y)^2} = 2N\cos^2(\vartheta) + 4N^2 \sin^2(\vartheta)$. 

We can now turn the attention to the final correlator $\braket{N^z N^y}$. For both the diagonal and off-diagonal terms we find 
$\braket{  \sigma^z_{j,A} \sigma^y_{k,A} - \sigma^z_{j,A}\sigma^y_{k,B} - \sigma^z_{j,B}\sigma^y_{k,A} + \sigma^z_{j,B} \sigma^y_{k,B} } = 2 \sin(2\vartheta) \sin(\vartheta)$. The total correlator therefore is  $\braket{N^yN^z} =2N^2 \sin(2\vartheta) \sin(\varphi) $. In total

\begin{align}
    \braket{(N^z)^2} &\to \braket{(N^z)^2} \cos^2(\vartheta_0) + 2\braket{N^zN^y} \cos(\vartheta_0) \sin(\vartheta_0) + \braket{(N^y)^2} \sin^2(\vartheta_0) \ . \label{eq:Nz2rotated}
\end{align}

\noindent We can substitute the rotated expectation value \eqref{eq:Nz2rotated} into the integral to find 

\begin{align}
    \braket{(N^z)^2}_\rho &= \frac 1 Z \int d\vartheta d \varphi 
    e^{-\frac{(1- \cos(\vartheta))^2}{2\sigma^2}} \ \braket{(N^z)^2}\\  \begin{split} &=
N \cos^2(\theta)  \left(-2 \sigma^2 + 4 N (1+\sigma^2) + \frac{4(1-2N)\sqrt{\frac{2}{\pi}}\sigma^{-1}\operatorname{erf}\left(\frac{\sqrt{2}}{\sigma}\right)}{\sigma}\right) \\
&+ N \left(2+(1-2 N) \sigma^2 + \frac{2(-1+2N)\sqrt{\frac{2}{\pi}}\sigma^{-1}\operatorname{erf}\left(\frac{\sqrt{2}}{\sigma}\right)}{\sigma}\right) \sin^2(\theta) \\
&+ \frac{N^2 \sqrt{2\pi} \left( {}_2F_1 \left(\frac{3}{4}, \frac{5}{4}; \frac{3}{2}, 2; -\frac{2}{2\sigma^2}\right) - {}_2F_1\left(\frac{5}{4}, \frac{7}{4}; 2, \frac{5}{2}; -\frac{2}{2\sigma^2}\right) \right) \sin(2\theta)}{\sigma \operatorname{erf}\left(\frac{\sqrt{2}}{\sigma}\right)} 
\end{split} \ .
\end{align}

\noindent Here, $\mathrm{erf}(x)$ denotes the error function whereas $F_n(a, b; c, d; g) $ denotes the generalized hypergeometric function. While this equation is rather complicated we can analyse some limits more closely. First in the limit $\sigma\to 0$ we recover 

\begin{align}
    \braket{(N^z)^2}_\rho = 4N^2 \cos^2(\vartheta_0) + 2N \sin^2(\vartheta_0) \ . 
\end{align}

\noindent The uncertainty $(\Delta N^z)^2 = 2N\sin^2(\vartheta) \propto N$ meaning there are no macroscopic quantum fluctuations in the state. Contrary in the limit $\sigma \to \infty$ we obtain 

\begin{align}
    \braket{(N^z)^2}_\rho = \frac 4 3 N(N+1) \ .
\end{align}

\noindent The uncertainty is given by $(\Delta N^z)^2 = \frac 4 3 N(N+1)\propto N^2$. This means in this case there are fluctuations scaling with system size. This means they stay finite even in the thermodynamic limit. If we take the limit $N\to \infty$ but keep $\sigma$ finite and set $\theta_0 = \pi/2$ we find 

\begin{align}
\sigma_{nz}^2 = 4\sigma \left( \pi \sigma - \frac{2 \left( \sqrt{2\pi} \right)}{\text{Erfi} \left( \frac{\sqrt{2}}{\sigma} \right)} \right) \ .
\end{align}

\noindent This means in the thermodynamic limit the fluctuations vanish $\sigma_{nz} \to 0$ for $\sigma\to 0$. For $\sigma \to \infty$ we find $\sigma_{nz}$ saturates to $\sigma_{nz}^2 \to 4/3$  consistent with the previous results. Crucially, for any finite $\sigma_{nz}>0$ value the fluctuations are finite.

\begin{figure}
    \centering
    \includegraphics[width=0.7\linewidth]{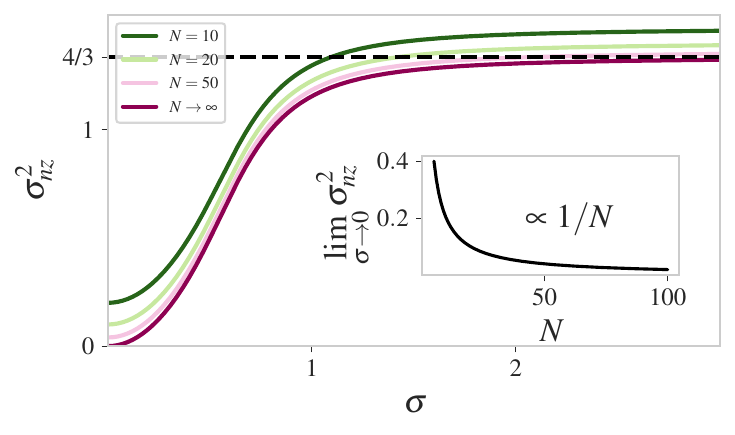}
    \caption{\justifying The figure shows $\sigma_{nz}^2$ as a function of $\sigma$ for different values of $N$ (violet to green). The function interpolates between the minima $2/N$ at $\sigma \to 0$ and saturates for $\sigma\to \infty, N\to \infty$ to $\sigma_{nz}^2 = 4/3$. The inset shows the minimum value of $\sigma_{nz}$ for $\sigma \to 0$. The minimum uncertainty is due to quantum fluctuations.}
    \label{fig:enter-label}
\end{figure}

\section{Dynamics of observables}

In this section, we will use the analytical form derived for the spectrum to analyze the equilibration dynamics of different observables. We will first investigate the evolution of \(\braket{n^z}(t)\) at $\alpha = 0$ and then turn our attention to the finite $\alpha>0$ case. 

\subsection{\texorpdfstring{Time Evolution of \(N^z\)}{Time Evolution of Nz}}

In this section, we derive the time evolution of the operator $N^z$ under certain assumptions motivated by numerical results. We consider mixed Néel states states (Eq. (2) in the main text) and make specific assumptions regarding the form of the initial state and the operator $N^z$ in the eigenbasis of the Hamiltonian.

As explained in the main text, we express the initial state $|\theta\rangle$ as a superposition of energy eigenstates $|n^z, s^z\rangle$

\begin{equation}
|\theta\rangle = \sum_{\substack{N^z, S^z\\ \ |N^z+ S^z| \leq N/2}} c_{N^z, S^z}(\theta) |N^z, S^z\rangle \ ,
\label{eq:initial_state}
\end{equation}
where the coefficients $c_{N^z, S^z}(\theta)$ are given by

\begin{equation}
c_{N^z, S^z}(\theta) = \frac{1}{2\pi\sigma_{sz}\sigma_{nz} N^2} \exp\left(- \frac{(S^z)^2}{2\sigma_{sz}^2N^2}\right) \exp\left(- \frac{(N^z - N^z_{\mathrm{stat}})^2}{2\sigma_{nz}^2N^2} \right).
\label{eq:coefficients}
\end{equation}

\noindent Here, $N^z$ and $S^z$ are indices associated with the eigenstates, $N^z_{\mathrm{stat}}$ is the center of the Gaussian distribution for $N^z$, and $\sigma_{sz}$ and $\sigma_{nz}$ are the standard deviations of the Gaussian distributions for $s^z$ and $n^z$, respectively.

We assume that the operator $N^z$ in the eigenbasis has the following form (see also main text) 

\begin{equation}
\mathcal N^z_{N_1^z, N_2^z} = N_1^z \delta_{N_1^z, N_2^z} + (N_0^z -N_1^z)  \delta_{N_1^z, N_2^z \pm 1},
\label{eq:Nz_operator}
\end{equation}
where $N_0^z$ is the initial condition. This form indicates that $N^z$ has diagonal elements proportional to $n^z$ and off-diagonal elements connecting states with $n^z$ differing by $\pm 1$. 
We highlight that the quantum operator $\mathcal N^z$ is strictly distinct from the index $N^z$ labeling the eigenfunctions. The operator $\mathcal N^z$ does not commute with the Hamiltonian and is therefore not diagonal in the eigenbasis.

We aim to calculate the time evolution of the expectation value of $N^z$, which is given by

\begin{equation}
\langle \theta | \mathcal N^z(t) | \theta \rangle = \langle \theta | e^{i\mathcal{H}t} \mathcal N^z e^{-i\mathcal{H}t} | \theta \rangle \ .
\end{equation}
Using the assumptions above, we can rewrite this as

\begin{equation}
\langle \theta | N^z(t) | \theta \rangle = \underbrace{\langle \theta | N_1^z \delta_{N_1^z, N_2^z} | \theta \rangle}_{\mathcal N_0} + \underbrace{\langle \theta | e^{i\mathcal{H}t}(N_0^z -N_1^z) \delta_{N_1^z, N_2^z \pm 1} e^{-i\mathcal{H}t} | \theta \rangle}_{\mathcal N_1^z(t)}.
\label{eq:Nz_time_evolution}
\end{equation}

\noindent We will now calculate $\mathcal{N}_0$, which represents the long-time average, as it is time-independent. $\mathcal{N}_1^z(t)$ represents the time-dependent part, which we will evaluate in the next section. \\ 

\begin{center}
    \textbf{Long-Time Averages $\mathcal{N}_0$}
\end{center}

Substituting the Ansatz for the initial state in the eigenbasis \eqref{eq:initial_state} and the assumed form of the $N^z$ operator \eqref{eq:coefficients} we can express $\mathcal N_0$ \eqref{eq:Nz_time_evolution} as

\begin{align}
\mathcal{N}_0^z/N&= \langle \theta | n^z \delta_{N_1^z,  N_2^z} | \theta \rangle \nonumber= \sum_{N_1^z, S_2^z}  \sum_{N_2^z, S_2^z} c_{N^z, S^z}^* c_{n_1^z, s_1^z} \langle n^z, s^z | n_1^z \delta_{N_1^z,  N_2^z}  | n_1^z, s_1^z \rangle \nonumber \\
&= \sum_{n^z}\frac 1 N  \frac{N^z}{N} \frac{\exp\left(- \frac{((N^z/N) - (N^z_{\mathrm{stat}}/N))^2}{2\sigma_{nz}^2}\right)}{\sqrt{2\pi}\sigma_{nz}^2 } \sum_{S^z=-N+ |n^z|}^{N-|N^z|} \frac 1 N \frac{\exp\left(- \frac{(S^z/N)^2}{2\sigma_{sz}^2}\right)}{\sqrt{2\pi}\sigma_{sz}^2}.
\label{eq:N0_initial}
\end{align}

\noindent We can now approximate the sums via an integral in the continuum limit. We identify the integration variables $n^z = N^z/N$ and $s^z = S^z/N$ finding

\begin{align}
     \mathcal N_0^z/N  = \int_{-1}^1 dn^z \ n^z \frac{e^{(n^z-n_{\mathrm{stat}}^z)^2/(2\sigma_{nz}^2)}}{{\sqrt{2\pi} \sigma_{sz}}} \ \int_{-1+ |n^z|}^{1 - |n^z|} ds^z \ \frac{\exp(-(s^z)^2 /(2\sigma_{sz}^2)}{\sqrt{2\pi} \sigma_{nz}} \ .
\end{align}

\noindent Assuming $\sigma_{sz}$ and $\sigma_{nz}$ are sufficiently narrow we can extend the two integrals up to infinity. The second integral over $s^z$ evaluates in this case to one. The first integral yields  

\begin{align}
    \mathcal N_0^z/N = n_{\mathrm{stat}}^z  \ .
\end{align} \\

\begin{center}
    \textbf{Time-Dependent Part: $\mathcal{N}_1^z(t)$}
\end{center}
We now turn our attention to the time-dependent part, $\mathcal{N}_1^z(t)$. Using \eqref{eq:Nz_time_evolution}, we can write $\mathcal{N}_1^z(t)$ explicitly as

\begin{align}
\mathcal N_1^z(t) &= \langle \theta | e^{i\mathcal{H}t}(N_0^z - N_1^z)  \delta_{N_1^z,  N_2^z \pm 2} e^{-i\mathcal{H}t} | \theta \rangle \nonumber \\
&= \sum_{N^z, S^z} |c_{N^z, S^z}|^2 \left[ (N_0^z - N^z-1) e^{i(E(N^z,S^z)-E(N^z+1, S^z))t} + (N_0^z + N^z+1) e^{i(E(N^z, S^z)-E(N^z-1, S^z))t} \right]. \label{eq:N1_initial}
\end{align}
In the last line, we have performed the sum over $n_1^z$, $s_1^z$ using the orthogonality of the eigenstates and the fact that $n_1^z = n^z \pm 1$. Furthermore, we used the definition of the eigenenergies $E(N^z/N, S^z/N)$ according to Eq. \eqref{eq:E_of_nz_sz}. 

In the following, we will drop terms of order $\mathcal{O}(1/N)$ in the energy difference in the exponent. This is justified for large enough $N$. Substituting \eqref{eq:coefficients} into \eqref{eq:N1_initial}, we obtain

\begin{align}
\mathcal{N}_1^z(t) &= \sum_{N^z, S^z} \frac{\exp\left(- \frac{(N^z - N^z_0)^2}{2\sigma_{nz}^2 N^2 }\right)}{\sqrt{2\pi\sigma_{nz}^2 N^2}} \frac{\exp\left(- \frac{(S^z)^2}{2\sigma_{sz}^2N^2}\right)}{\sqrt{2\pi \sigma_{sz}^2 N^2}} \left[ (N_0^z - N^z-1) e^{-i(E((N^z+1)/N,S^z/N)-E(N^z/N, S^z/N))t} \right. \nonumber \\
&\quad \left. + (N_0^z + N^z+1) e^{-i(E((N^z-1)/N, S^z/N)-E(N^z/N, S^z/N))t} \right].
\label{eq:N1_simplified}
\end{align}
Let us define the energy difference $\Delta E = E(n^z \pm 1/ N, s^z) - E(n^z, s^z)$. For simplicity we write $E(n^z, s^z, N) = h_N N n^z + JN(s^z)^2 + N f(n^z, s^z)$ where $f(n^z, s^z) = -\sqrt{(h_N + Jn^z)^2 + \frac{J^2}4 (1 + (n^z)^2 -(s^z)^2 )^2}$.  

\begin{align}
\Delta E &= E(n^z + 1/N, s^z) - E(n^z, s^z) = h_N + N( f(n^z + \frac 1 N, s^z) - f(n^z, s^z))\ .
\label{eq:Delta_E}
\end{align}

\noindent We will now replace $x = 1/N$. Then we find for large $N$ we can approximate the difference as a derivative

\begin{align}
    \Delta E &=  h_N +\bigg( \frac{f(n^z+x)- f(n^z)}{x}\bigg) \approx  h_N +  \frac{df}{dn^z} \\ &= h_N  +  \frac{2J(h_N +Jn^z) + J^2 n^z ( 1 + (n^z)^2 - (s^z)^2)   }{f(n^z, s^z)} \ .
\end{align}

\noindent We now expand this expression around $n^z \approx n_{\mathrm{stat}}^z$ and $s^z \approx 0$ which is valid if $\sigma_{nz}, \sigma_{sz} \ll 1$ 

\begin{equation}
\lim_{N\to \infty} \Delta E \approx \omega_p + \gamma_{n^z} (n^z - n_0^{\mathrm{stat}}) + \gamma_{sz}(s^z)^2  .
\label{eq:Delta_E_expansion}
\end{equation}

\noindent The first term $\omega_p$ is the pendulum frequency. The expression is generally quite complicated. Here we show it in the simplifying case $n_{\mathrm{stat}}^z\approx -1$

\begin{equation}
\omega_p = h_N + \frac{J \left(  2J-h_N\right)}{\sqrt{(h_N -J)^2 + J^2}}
\ .
\label{eq:omega_p}
\end{equation}

\noindent We note in the case $J\approx h_N$ this recovers to the mean-field result $\omega_p \propto \sqrt{J h_N}$. $\gamma_{nz}$ and $\gamma_{sz}$ on the other hand are (again for $n_{\mathrm{stat}}^z \approx -1$) given by 

\begin{align}
\gamma_{nz} &= 2J^2 \frac{(h_N - J)^2}{\left((h_N - J)^2 + J^2\right)^{3/2}} \label{eq:gamma_nz}, \\
\gamma_{sz} &= \frac{h_N\,J^2\,(J-h_N)}{2\left(h_N^2-2h_NJ+2J^2\right)^{3/2}} \label{eq:gamma_sz}.
\end{align}

\noindent Furthermore, we will approximate $n^z - n^z \approx n_0^z - n_{\mathrm{stat}}^z$. Substituting \eqref{eq:Delta_E_expansion} into \eqref{eq:N1_simplified}, we get

\begin{align}
\mathcal N_1^z(t) &\approx \sum_{s^z, n^z} \   \frac{\exp\left(- \frac{(n^z - n^z_0)^2}{2\sigma_{nz}^2}\right)}{\sqrt{2\pi \sigma_{nz}^2 N^2}} \frac{\exp\left(- \frac{(s^z)^2}{2\sigma_{sz}^2}\right)}{\sqrt{2\pi \sigma_{sz}^2 N^2}} \ \  (n_0^z - n^z) 
  \cos\left[\left(\omega_p + \gamma_{nz} \left(n^z + 1\right) + \gamma_{sz} (s^z)^2 \right)t\right].
\label{eq:N1_approx}
\end{align}

\noindent We can now separate the sums over $n^z$ and $s^z$ to analyze the different time scales (discussed in the main text)

\begin{equation}
\boxed{\mathcal N_1^z(t) =  \mathcal N^z_{\mathrm{pend}}(t) \mathcal N^z_{\mathrm{comb}}(t) \mathcal N^z_{\mathrm{eq}}(t)} \ ,
\label{eq:N1_separation}
\end{equation}
\noindent where we have defined 

\begin{align}
\mathcal N^z_{\mathrm{pend}}(t) &=  (n_0^z - n^z) \cos\left[2\left(\omega_p + \gamma_{nz}\left(2n_0^z+ \frac{h_N}{J}\right) \right)t\right] \label{eq:npend} \ ,\\ 
\mathcal N_{\mathrm{comb}}^z(t) &= \mathrm{Re\bigg[ } \sum_{n^z}  \frac{\exp\left(- \frac{(n^z - n^z_0)^2}{2\sigma_{nz}^2}\right)}{\sqrt{2\pi \sigma_{nz}^2}} e^{2i\gamma_{nz} (n^z- n_0^z)t }\bigg] ,
\label{eq:N1_n}\\ 
\mathcal N_{\mathrm{eq}}^z(t) &= \mathrm{Re}\bigg[\sum_{s^z} \frac{\exp\left(- \frac{(s^z)^2}{2\sigma_{sz}^2}\right)}{\sqrt{2\pi \sigma_{sz}^2}} e^{2i\gamma_{sz} (s^z)^2 t}\bigg]\ .
\label{eq:N1_s}
\end{align}

\noindent The above equations describe the decomposition of the time evolution of $N^z$ into three components, each associated with a distinct phase discussed in the main text. Equation \eqref{eq:npend} for $\mathcal N^z_{\mathrm{pend}}(t)$ captures the initial pendulum-like oscillations, driven by the effective frequency $\omega_p$ and modified by $\gamma_{nz}$, $n_0^z$, and the ratio $h_N/J$. This term dominates the short-time dynamics. Equation \eqref{eq:N1_n} will give rise to equilibration in intervals while \eqref{eq:N1_s} will give rise to equilibration on average. We will discuss these terms in greater detail in the following.  \\ 

\begin{center}
    \textbf{Time Evolution Analysis of $\mathcal{N}_{\mathrm{comb}}^z(t)$}
\end{center}

This section details the derivation of $\mathcal{N}_{\text{comb}}^z(t)$, which describes the "equilibration in intervals" phase shown in the second panel of Fig.~1 (c) in the main text. We begin with equation \eqref{eq:N1_n} 

\begin{equation}
\mathcal N_{\mathrm{comb}}^z(t) = \mathrm{Re}\bigg[ \sum_{N^z= -N}^N  \frac 1 N  \frac{\exp\left(- \frac{(N^z - N^z_{\mathrm{stat}})^2}{2\sigma_{nz}^2 N^2 }\right)}{\sqrt{2\pi \sigma_{nz}^2}} e^{i\gamma_{nz}(\frac{N^z}N - \frac{N^z_{\mathrm{stat}}}{N}) t }\bigg].
\label{eq:N_comb_fourier_supp_B}
\end{equation}

\noindent  This expression has the form of a discrete Fourier series with equally spaced frequencies $\omega_{N^z} = \gamma_{nz} \frac{N^z}{N}$ with $N^z \in \{-N, -N+1, \dots, N-1, N\}$. Therefore, it is periodic with a revival time of (see also Fig. 2 (e) of the main text)

\begin{equation}
\boxed{T_{\mathrm{rev}} = \frac{2\pi N}{\gamma_{nz}}\ . }
\label{eq:T_revival_comb_supp_B}
\end{equation}

\noindent In the large $N$ limit, we can replace $N^z/N$ by the continuous variable $n^z = N^z/N$. The sum  \eqref{eq:N_comb_fourier_supp_B} then becomes a Riemann integral $n^z$

\begin{equation}
\mathcal{N}_\text{comb}^z(t) \approx \text{Re}\left[\frac{1}{\sqrt{2\pi\sigma_{nz}^2}} \int_{-\infty}^{\infty} dn^z \exp\left(- \frac{(n^z - n^z_{\mathrm{stat}})^2}{2\sigma_{nz}^2} + i \gamma_{nz} (n^z - n_{\mathrm{stat}}^z) t\right)\right].
\label{eq:N_comb_integral_supp_B}
\end{equation}

\noindent Evaluating the Gaussian integral in \eqref{eq:N_comb_integral_supp_B} yields

\begin{equation}
\mathcal{N}_\text{comb}^z(t) = \sum_{n=-\infty}^{\infty} G(t - nT_{\text{revival}})\ ,
\label{eq:N_comb_G_supp_B}
\end{equation}

\noindent where $G(t)$ denotes the shape of the revivals

\begin{equation}
G(t) = \exp\left(-\frac{t^2}{2T_{\text{decay}}^2}\right) 
\label{eq:G_comb_supp_B} \ .
\end{equation}

\noindent Here, we have defined the decay time also given in the main text 

\begin{equation}
\boxed{T_{\mathrm{dec}} = \frac{1}{\gamma_{nz} \sigma_{nz}}
\label{eq:T_decay_comb_supp_B} \ .}
\end{equation}

Equation \eqref{eq:N_comb_G_supp_B} represents $\mathcal{N}_\text{comb}^z(t)$ as a sum of revivals centered at $t = nT_{\text{revival}}$, each with a shape described by $G(t)$.  $G(t)$, defined in \eqref{eq:G_comb_supp_B}, consists of a Gaussian envelope with a width determined by the decay time $T_{\text{decay}}$. These parameters, along with the revival time $T_{\text{rev}}$, characterize the ``equilibration in intervals" phase, where the system exhibits oscillations that decay and revive periodically, reflecting a partial equilibration within each revival period. In the main text these dynamics are shown in the time domain in Fig.~1 (c) and in the frequency domain in Fig.~2 (e). \\

\begin{center}
    \textbf{Long-Time Evolution according to $\mathcal N_{\mathrm{eq}}^z(t)$}
\end{center}

This section details the derivation of $\mathcal{N}_{\text{eq}}^z(t)$, which describes the long-time "equilibration on average" phase. The dynamics are shown in the third panel of Fig.~1(c) in the main text.  We begin with the expression

\begin{equation}
\mathcal N_{\mathrm{eq}}^z(t) = \mathrm{Re}\bigg[\sum_{S^z}\frac 1 N \frac{\exp\left(- \frac{1}{2\sigma_{sz}^2}\left(\frac{S^z}{N}\right)^2\right)}{\sqrt{2\pi \sigma_{sz}^2}} e^{i\gamma_{sz} (S^z/N)^2 t}\bigg].
\label{eq:N_eq_start_supp_D}
\end{equation}

In the large $N$ limit, and for sufficiently long times, we can approximate the discrete sum in \eqref{eq:N_eq_start_supp_D} by an integral over $s^z$

\begin{equation}
\mathcal N_{\mathrm{eq}}^z(t) \approx \mathrm{Re}\left[\frac{1}{\sqrt{2\pi \sigma_{sz}^2}} \int_{-\infty}^{\infty} ds^z \exp\left(- \frac{(s^z)^2}{2\sigma_{sz}^2} + i\gamma_{sz} (s^z)^2 t\right)\right].
\label{eq:N_eq_integral_supp_D}
\end{equation}

\noindent This is a Gaussian integral, which evaluates to

\begin{equation}
\mathcal N_{\mathrm{eq}}^z(t) \approx \mathrm{Re}\left[\frac{1}{\sqrt{1 - 2i\gamma_{sz}\sigma_{sz}^2 t}}\right].
\label{eq:N_eq_intermediate_supp_D}
\end{equation}

\noindent Simplifying this expression by writing the complex number in polar form, we find

\begin{equation}
\mathcal N_{\mathrm{eq}}^z(t) = \frac{\cos\left(\frac{1}{2}\arctan(4\gamma_{sz}\sigma_{sz}^2 t)\right)}{\sqrt[4]{1 + (4\gamma_{sz}\sigma_{sz}^2 t)^2}}\ .
\label{eq:N_eq_final_supp_D}
\end{equation}

For long times, $t \gg \frac{1}{4\gamma_{sz}\sigma_{sz}^2}$, the denominator of \eqref{eq:N_eq_final_supp_D} is dominated by the term $(4\gamma_{sz}\sigma_{sz}^2 t)^2$, leading to a power-law decay

\begin{equation}
|\mathcal N_{\mathrm{eq}}^z(t)| \approx \frac{1}{2\gamma_{sz}\sigma_{sz}^2 t}\ .
\end{equation}

\noindent Thus, we identify the decay time for this long-time equilibration as (see also the main text)

\begin{equation}
\boxed{ T_{\mathrm{eq}} = \frac{1}{4\gamma_{sz}\sigma_{sz}^2}\ . }
\label{eq:T_decay_eq_supp_D}
\end{equation}

\noindent Equation \eqref{eq:N_eq_final_supp_D} describes the ``equilibration on average" phase, characterized by a slow power-law decay with a time scale $T_{\text{eq}}$ as shown in the main text in Fig.~1 (c) and Fig.~2(e).  In this long-time limit, the system approaches a steady state, and the absence of revivals ($T_{\text{rev}} \to \infty$) indicates that a true equilibrium is reached.

\subsection{Time evolution of the reduced density matrix}

This section investigates the behavior of the reduced density matrix, obtained by tracing out the $S^z$ degree of freedom. This is relevant for observables that do not depend strongly on $S^z$, such as the Néel parameter and generalizes the results discussed in the main text.
The time-dependent density matrix is given by

\begin{equation}
\rho(t) =e^{-i\mathcal{H}t} |\psi\rangle \langle\psi| e^{i\mathcal{H}t}= \sum_{n^z, s^z} |c_{n^z, s^z}|^2 |n^z, s^z\rangle \langle n^z, s^z| + \sum_{\substack{n_1^z\neq n_2^z \\ s_1^z \neq s_2^z}} c_{n_1^z, s_1^z}c^*_{n_2^z, s_2^z} e^{i(E(n_1^z, s_1^z) - E(n_2^z, s_2^z))t} |n_2^z, s_2^z\rangle\langle n_1^z, s_1^z|\ .
\label{eq:rho_decomposition_C}
\end{equation}

\noindent where $\ket\psi$ is the initial state, $c_{n^z, s^z}$ are the expansion coefficients of the initial state in the eigenbasis \eqref{eq:coefficients}, and $E(n^z, s^z)$ are the energy eigenvalues given by \eqref{eq:energy_difference_expansion_C}.
For our choice of initial state (equation \eqref{eq:coefficients}), the coefficients factorize as

\begin{equation}
|c_{n^z, s^z}|^2 = |c_{n^z}|^2 \frac{e^{-(s^z)^2/(2\sigma_{sz}^2)}}{\sqrt{2\pi \sigma_{sz}^2}N}\ .
\label{eq:c_factorization_C}
\end{equation}

\noindent The reduced density matrix $\mathrm{Tr}_{S^z} (\rho(t))$ is obtained by tracing out the $S^z$ degree of freedom

\begin{equation}
\mathrm{Tr}_{S^z}(\rho(t))= \sum_{s^z}\braket{s^z|\rho(t)|s^z}.
\label{eq:reduced_rho_def_C}
\end{equation}

\noindent The diagonal elements are given by

\begin{equation}
[\mathrm{Tr}_{S^z}(\rho(t))]_{n^z, n^z}  = |c_{n^z}|^2 \sum_{s^z} \frac{e^{-(s^z)^2/(2\sigma_{sz}^2)}}{\sqrt{2\pi \sigma_{sz}^2}N} = |c_{n^z}|^2,
\label{eq:reduced_rho_diagonal_C}
\end{equation}

\noindent where we used the normalization condition $\sum_{s^z} \frac{e^{-(s^z)^2/(2\sigma_{sz}^2)}}{\sqrt{2\pi \sigma_{sz}^2}N} = 1$. For the off-diagonal elements ($n_1^z \neq n_2^z$), we have

\begin{equation}
[\tilde{\rho}(t)]_{n_1^z, n_2^z} = \sum_{s^z} [\rho(t)]_{n_1^z, n_2^z, s^z, s^z} = \left(\sum_{s^z} \frac{e^{-(s^z)^2/(2\sigma_{sz}^2)}}{\sqrt{2\pi \sigma_{sz}^2 N^2}} e^{i(E(n_1^z, s^z) - E(n_2^z, s^z))t}\right) c_{n_1^z}c^*_{n_2^z}.
\label{eq:reduced_rho_offdiagonal_C}
\end{equation}

\noindent Assuming $\sigma_{sz}\ll  1$, we expand the energy difference to second order in $s^z$ around $s^z = 0$

\begin{equation}
E(n_1^z, s^z) - E(n_2^z, s^z) \approx E(n_1^z, 0) - E(n_2^z, 0) + \gamma(n_1^z, n_2^z) (s^z)^2,
\label{eq:energy_difference_expansion_C}
\end{equation}

\noindent where

\begin{align}
\gamma(n_1^z, n_2^z) &= N( \eta(n_1^z)
-\eta(n_2^z)), \\ \eta(n^z) &= \frac{J^2 \Bigl(1 + \bigl(n^z\bigr)^2\Bigr)}
{4\,\sqrt{\left(h_N + J\,n^z\right)^2 + \dfrac{J^2}{4}\Bigl(1 + \bigl(n^z\bigr)^2\Bigr)^2}}
\label{eq:gamma_n1_n2_C}
\end{align}

\noindent We highlight that this expression seems to scale with $N$. However, usually one will consider matrix elements close in $n^z$ meaning $n_2^z = n_1^z + \Delta n/N$ where $\Delta n$ is some fixed integer. In this case $\gamma(n_1^z, n_2^z)= N(\eta(n^z + \Delta n/N) -\eta(n^z)) = \Delta n\frac{d\eta}{dn^z} \propto \mathcal O(1)$.  Substituting \eqref{eq:energy_difference_expansion_C} into \eqref{eq:reduced_rho_offdiagonal_C}, we obtain

\begin{equation}
[\tilde{\rho}(t)]_{n_1^z, n_2^z} = f_{n_1^z, n_2^z}(t) c_{n_1^z}c^*_{n_2^z} e^{i(E(n_1^z) - E(n_2^z))t},
\label{eq:reduced_rho_offdiagonal_2_C}
\end{equation}

\noindent where

\begin{equation}
f_{n_1^z, n_2^z}(t) = \sum_{s^z} \frac{e^{-(s^z)^2/(2\sigma_{sz}^2)}}{\sqrt{2\pi \sigma_{sz}^2}N} e^{i\gamma(n_1^z, n_2^z)(s^z)^2 t}.
\label{eq:f_n1_n2_C}
\end{equation}

\noindent The expression for $f_{n_1^z, n_2^z}(t)$ describes a time-dependent factor that modulates the off-diagonal elements of the reduced density matrix. It has a recurrence time of

\begin{equation}
T_{\mathrm{rec}} = \frac{ 2\pi N^2}{\gamma(n_1^z, n_2^z)}.
\label{eq:recurrence_time_C}
\end{equation}

\noindent We note this is for specific $n_1^z$ and $n_2^z$. For the total operator the revival times can be significantly longer. In the thermodynamic limit ($N \to \infty$), we expect this recurrence time to diverge. In this limit, and for times much smaller than the recurrence time we can approximate the sum in \eqref{eq:f_n1_n2_C} by an integral, yielding

\begin{equation}
|f_{n_1^z, n_2^z}(t)| \approx \frac{1}{\sqrt[4]{1 + (t/T_{\mathrm{decay}})^2}}\ ,
\label{eq:f_n1_n2_modulus_C}
\end{equation}

\noindent where we have defined $T_{\mathrm{decay}} = 1/(\gamma(n_1^z, n_2^z) \sigma_{sz}^2)$. This factor determines the decay of the off-diagonal elements. We can now write the entire reduced density matrix as

\begin{equation}
\tilde \rho(t) =\sum_{n^z} |c_{n^z}|^2 |n^z\rangle \langle n^z| + \sum_{n_1^z\neq n_2^z}  f_{n_1^z, n_2^z}(t) c_{n_1^z }c^*_{n_2^z} \ e^{i(E(n_1^z)- E(n_2^z))t} |n_1^z\rangle\langle n_2^z|\ .
\label{eq:reduced_rho_final_C}
\end{equation}

The behavior of $\tilde{\rho}(t)$ depends on the scaling of the fluctuations in $s^z$ with system size $N$. We note $\gamma(n_1^z, n_2^z)$ does not scale with $N$ if the state is sufficiently localized in the eigenbasis meaning $|c_{n_1^z} c^*_{n_2^z}| \approx 0$ if $|n_1^z - n_2^z| \gg 0$. 

If the variance $\sigma_{sz}^2$ stays constant as the system size increases (macroscopic fluctuations), the decay time $T_{\mathrm{decay}} \propto \mathcal O(1)$ is system size independent.  As a result, $f_{n_1^z, n_2^z}(t)$ decays to zero for any finite time $t$ in the thermodynamic limit. Consequently, the reduced density matrix equilibrates to the diagonal ensemble

\begin{equation}
\lim_{t\to \infty} \tilde \rho(t) = \sum_{n^z} |c_{n^z}|^2 |n^z\rangle \langle n^z|\ .
\label{eq:reduced_rho_case1_C}
\end{equation}

If, however, $\sigma_{sz}^2$ scales as $\sigma_{sz}^2 \propto 1 / N$ (for $\sigma = 0$), the decoherence time is given by (result presented in the main text)

\begin{equation}
\boxed{T_{\mathrm{eq}} = \frac{1}{\gamma(n_1^z, n_2^z)\sigma_{sz}^2} = \frac{N}{\gamma(n_1^z, n_2^z) }\ . }
\label{eq:decoherence_time_case2_C}
\end{equation}

\noindent $T_{\text{eq}}$ diverges as $N^2$ in the thermodynamic limit. Consequently, the off-diagonal elements do not decay to zero, and the system does not equilibrate to the diagonal ensemble. In this case the mean field dynamics determine the full time evolution of the system.


\section{\texorpdfstring{Dynamics at finite $\alpha > 0$}{Dynamics at finite alpha > 0}}

In this section, will explore how the dynamics in the finite $\alpha > 0$ case differ from those in the $\alpha = 0$ case mainly discussed in the main text.

To assess the dynamics, we perform exact diagonalization of the full Hamiltonian [Eq. (1) in the main text]. This method allows only small system sizes to be studied; however, we can apply finite size scaling to extrapolate to larger systems. Additionally, we utilize two symmetries to reduce the Hilbert space dimension. First, we exploit the conservation of the total spin-z component, $S^z = \sum_j \sigma_{j,A}^z + \sigma_{j,B}^z$. Second, we impose translational invariance under shifts by two lattice sites, corresponding to the unit cell of the bipartite (A–B) structure.

To probe the spin dynamics, we choose an initial density matrix
\begin{align}
\rho(\theta, \phi) = \prod_j 
\int d\Omega_j \,\exp\left[-\frac{1 - (-1)^j \,\mathbf{r}(\Omega_j)\cdot \mathbf{r}(\Omega_0)}{2\,\sigma^2}\right]
\,\ket{\theta_j,\phi_j}\!\bra{\theta_j,\phi_j}, \label{eq:densityspinbased}
\end{align}
where $\Omega_j = (\theta_j, \phi_j)$ and $\Omega_0 = (\theta_0, \phi_0)$ set the orientation of a Néel‐type order parameter. In the limit $\sigma \to 0$, this state becomes the pure (rotated) Néel state $\ket{\theta_0}$. At finite $\sigma > 0$, it acquires small fluctuations in each local spin orientation, inducing fluctuations in the collective spin operators $\boldsymbol{S}_A^2$ and $\boldsymbol{S}_B^2$. Similar to the main text, this approach helps distinguish effects that are dependent on the initial state from those that depend on the spectrum.

At $\alpha = 0$, the length of the Néel parameter $\boldsymbol{N}^2 = (\boldsymbol{S}_A - \boldsymbol{S}_B)^2$ and the total spin length $\boldsymbol{S}^2 = (\boldsymbol{S}_A + \boldsymbol{S}_B)^2$ change over time. However, the lengths of the collective spins on the $A$ and $B$ sublattices, $\boldsymbol{S}_A^2$ and $\boldsymbol{S}_B^2$, remain constant. Consequently, the Hamiltonian governs the dynamics of two well-defined collective spins, while the degrees of freedom within each collective spin remain frozen.

\begin{figure}[ht]
    \centering
    \begin{overpic}[width=1.0\textwidth]{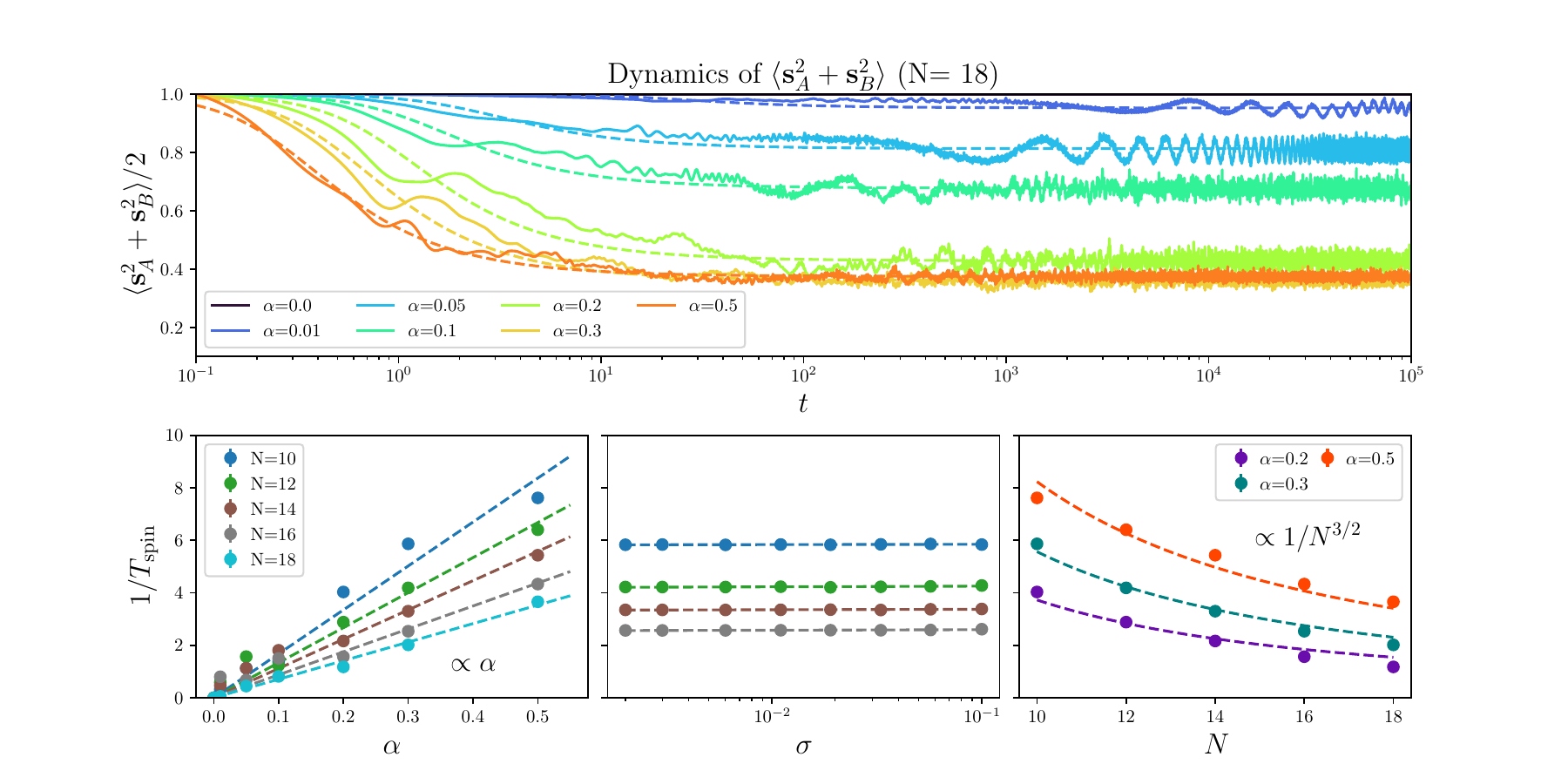} 
\put(7.1, 43){{ (a)}}
\put(7.4, 20.3){{(b)}}
\put(39, 20.3){{(c)}}
\put(66, 20.3){{(d)}}
\end{overpic}
    \caption{\justifying \small
(a)~Time evolution of the total sublattice spin length $\langle \boldsymbol s_A^2 + \boldsymbol s_B^2\rangle /2$ at $N=18$ for various $\alpha>0$, obtained via exact diagonalization of Eq.~(1) in the main text. Unlike the $\alpha=0$ case—where sublattice spins remain frozen—these collective spins decay to a stationary value below one, with persistent finite‐size oscillations at small $\alpha$. (b–d)~Decay time $T_{\mathrm{spin}}$ versus $\alpha$, initial state fluctuation $\sigma$, and system size $N$. In (b), $1/T_{\mathrm{spin}}$ increases linearly with $\alpha$, indicating faster decay for larger $\alpha$ and slower decay at larger $N$. Panel (c) demonstrates the insensitivity of $T_{\mathrm{spin}}$ to $\sigma$. Finally, (d) shows $T_{\mathrm{spin}}\propto N^{3/2}$, highlighting the pronounced finite‐size effects that prolong collective spin relaxation for larger $N$. The simulation parameters are $J/h = 2.5$ and $\theta_0 = \pi/2$. }
    \label{fig:spindecay}
\end{figure}

At finite $\alpha > 0$, the collective spin lengths $\boldsymbol{S}_A^2$ and $\boldsymbol{S}_B^2$ become dynamic variables as they no longer commute with the Hamiltonian. We investigate their dynamics using exact diagonalization, as shown in Figure 1(a). For any $\alpha > 0$, the total spin length $\boldsymbol{S}_A^2 + \boldsymbol{S}_B^2$ decays over time and reaches a stationary value below one, i.e., $(\boldsymbol{S}_A^2 + \boldsymbol{S}_B^2) < 1$. As $\alpha$ increases, the spin length decays more rapidly. 

Notably, the spin length does not completely decay to zero; however, this can be attributed to finite size effects and is expected to vanish for larger system sizes. The finite size effects are more pronounced at smaller $\alpha$ as the interaction tails cannot decay fully. While for larger $\alpha > 1$ the system reaches a long-time stationary state, at smaller $\alpha$ values such as $\alpha = 0.01$ and $0.05$, large oscillations persist at long times. This persistence is also a finite size effect, as the interaction tails do not fully decay within the system's bounds.

We further investigates the scaling of the spin decay time $T_{\mathrm{spin}}$ with different parameters $\alpha$, fluctuations in the initial state \eqref{eq:densityspinbased} $\sigma$ and system sizes $N$. As the system is close to integrable the level spacings follow a Possoinian distribution. Then the decay is approximately described by the power law 

\begin{align}
    f(t) = \frac 1 {1 + \left( t /{T_{\mathrm{spin}}}\right)^2} \ \ .
\end{align}

\noindent We extract $T_{\mathrm{spin}}$ by performing a fit with this equation [see dashed line in Fig. ~\ref{fig:spindecay} (a)].

Figure~\ref{fig:spindecay} (b), (c), and (d) illustrate the scaling of decay time $T_{\mathrm{spin}}$ with parameters $\alpha$, $\sigma$, and $N$. We find the following

\begin{enumerate}
\item Panel (b) shows $T_{\mathrm{spin}} \propto 1/\alpha$, evidenced by the linear increase of the decay rate $1/T_{\mathrm{spin}}$ with $\alpha$.  Slower decay is observed for larger system sizes ($N$), as indicated by the decreased slope of this linear trend. This underscores the strong inverse relationship between $T_{\mathrm{spin}}$ and $\alpha$, and the amplification of finite-size effects with increasing $N$.

\item Panel (c) demonstrates the invariance of $T_{\mathrm{spin}}$ to initial state fluctuations $\sigma$. The consistent decay time $T_{\mathrm{spin}}$ across varying $\sigma$ values suggests robustness to $\sigma$, implying its negligible influence on decay behavior.

\item Panel (d) reveals $T_{\mathrm{spin}} \propto N^{3/2}$. Larger systems exhibit prolonged decay times, evidenced by the monotonic decrease of $1/T_{\mathrm{spin}}$ with $N$. The $\alpha$ scaling is also apparent: smaller $\alpha$ corresponds to slower decay, and larger $\alpha$ to faster decay, highlighting the prominent role of finite-size effects on long-time dynamics.
\end{enumerate}

In the finite $\alpha$ case and at finite system sizes, the equilibration dynamics acquire a third stage—the decay of the collective spins in the $A$ and $B$ sublattices. Since the lifetime of the collective spin ($\propto N^{3/2}$) is asymptotically longer than the equilibration time of the Néel parameter (for states with minimal fluctuations, $\propto N$), the dynamics consisting of mean-field behavior, equilibration in intervals, and equilibration on average, as discussed in the main text, precede this decay. This implies that the long-wavelength degrees of freedom equilibrate faster than the short-ranged degrees of freedom.

In the thermodynamic limit, the behavior of the collective spins $\boldsymbol{S}_A^2$ and $\boldsymbol{S}_B^2$ differs significantly from that of the Néel parameter. Crucially, the lifetime of the collective spins depends directly on the system size, not on fluctuations in the initial state. While the Néel parameter can equilibrate even in the thermodynamic limit, finite-size scaling implies that the lifetime of the collective spins diverges irrespective of the initial state. Consequently, in the thermodynamic limit, the behavior for $0 < \alpha < 1$ is identical to that at $\alpha = 0$ justifying the approach in the main text to mainly discuss the $\alpha = 0$ case. 

\begin{figure}
    \centering
    \includegraphics[width=0.9\linewidth]{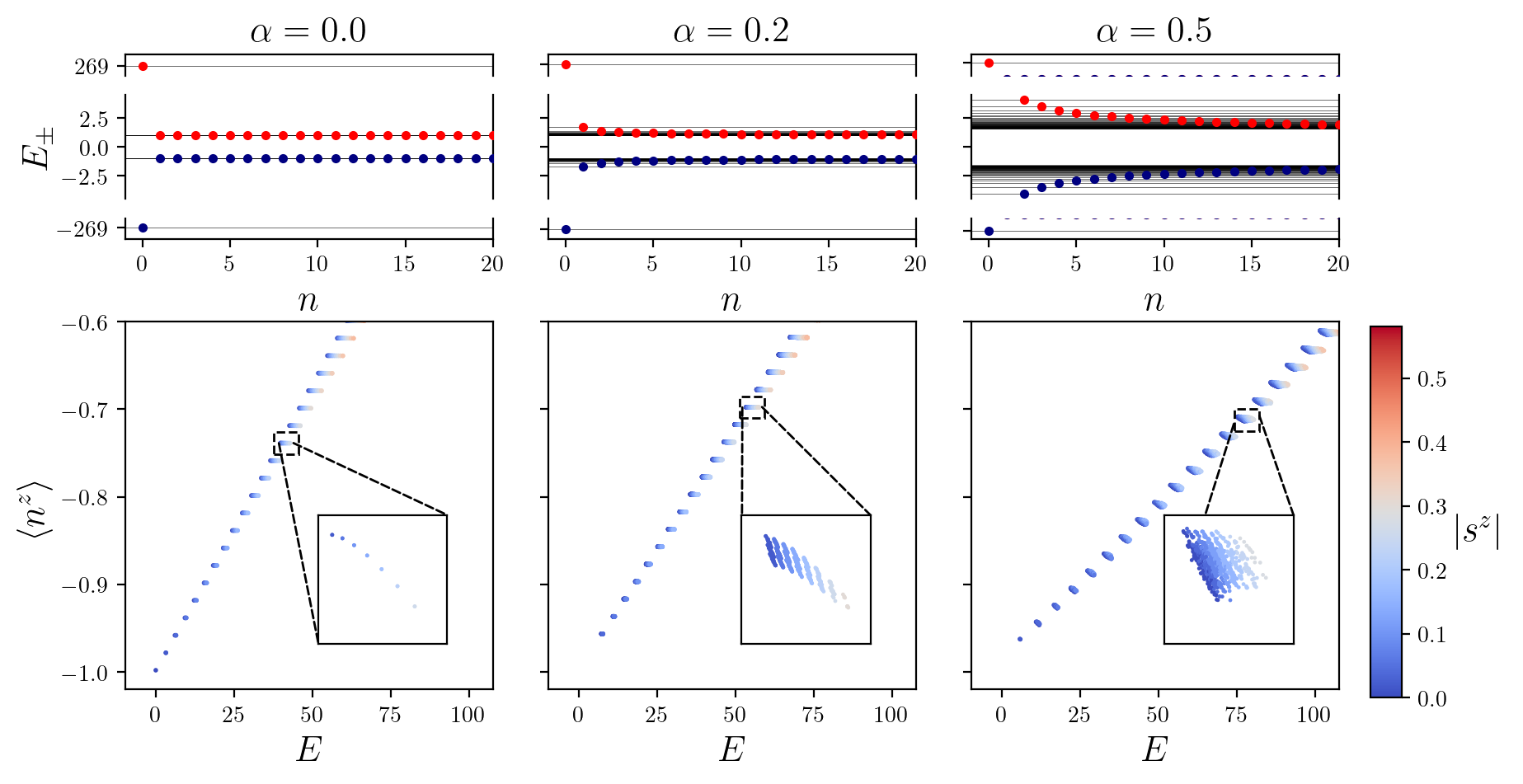}
    \caption{ \justifying\textit{Top row:} Single‐particle energies \(E_{A,n}\) (red disks) and \(E_{B,n}\) (blue disks) plotted against the mode index \(n\) for three different values of \(\alpha\).  At \(\alpha=0\) (left panel), all levels with \(n>2\) are degenerate; turning on \(\alpha>0\) (middle and right panels) lifts this degeneracy.  
\textit{Bottom row:} Many‐body energies \(E\) \eqref{eq:manybodyiniteAlpha} (horizontal axis) versus the self‐consistently determined \(\langle n^z\rangle\) \eqref{eq:nz} (vertical axis).  Each point corresponds to a many‐body eigenstate, and is colored by the spin expectation value \(\lvert s^z\rvert\) (blue to red).  The insets zoom in on one of the energy “clusters” that form when \(\alpha>0\), reflecting the splitting of the originally degenerate subspaces and leading to an additional (long) decay timescale for the collective spin.}
    \label{fig:finiteAlphaSpectrum}
\end{figure}

The system-size dependence of spin decay can be clarified by examining the analytical single-particle spectrum outlined in the main text. In the top row of Fig.~\ref{fig:finiteAlphaSpectrum}, we present the single-particle energies $E_{A/B,n} = \pm \sqrt{h_{N,n}^2 + \Gamma_n^2},$ which, at $\alpha = 0$, exhibit full degeneracy for all levels with $n>2$. For finite $\alpha > 0$, this degeneracy is lifted. In the regime of large $n$, the energies can be approximated by
\begin{align}
    E_{A/B,n} &\approx  \pm \Bigl(h_{N,n} 
    - \frac{\tilde J}{n^{\alpha -1}} \Bigr),
\end{align}
with $\tilde J = J\,n^z\,c_\alpha\,\Gamma(1-\alpha)\,\cos\bigl(\pi\alpha/2\bigr)$ showing that the levels approach accumulation points at $\pm h_N$ with a characteristic $1/n^{1-\alpha}$ scaling. Despite remaining discrete, the levels cluster around these accumulation points, and this clustering will play a crucial role in determining the many-body spectrum.
We now construct the many-body spectrum for finite $\alpha > 0$ to understand how the single-particle degeneracies affect the collective behavior. Each many-body state is specified by a set of occupied single-particle levels, and its energy is given by
\begin{align}
    E \;=\; \sum_n \Bigl(E_{A,n}\,\langle d_{A,n}^\dagger\,d_{A,n}\rangle 
            + E_{B,n}\,\langle d_{B,n}^\dagger\,d_{B,n}\rangle\Bigr)\ . \label{eq:manybodyiniteAlpha}
\end{align}
Here, $\langle n^z\rangle$ must be determined self-consistently. As illustrated in the bottom row of Fig.~\ref{fig:finiteAlphaSpectrum}, the $N^2$ degenerate subspaces at $\alpha=0$ split into multiplets containing exponentially many states that cluster near the original energy levels. The finite width of these clusters introduces an additional timescale, which determines the decay time of the collective spin, $T_{\mathrm{spin}}$.

We will now show that the width of each cluster effectively vanishes in the thermodynamic limit, explaining the divergence of the decay time for the collective spins. Each cluster can be defined by the occupancies in $ A $ and $ B $ with $ N_A = \sum_n \braket{d_{A,n}^\dagger d_{A,n}} $ and $ N_B = \sum_n \braket{d_{B,n}^\dagger d_{B,n}} $. Let us consider the level spacing between two randomly occupied states under the constraint $ N_A, N_B $. Under this constraint, different states can only differ by flips $ d_{n,A}^\dagger d_{m,A} $ or $ d_{n,B}^\dagger d_{m,B} $. Each of these flips contributes to the energy

\begin{align}
\Delta_{nm} = \tilde J \bigg( \frac{1}{n^{\alpha -1}} - \frac{1}{m^{\alpha -1}} \bigg)\ .
\end{align}

\noindent This means the energy difference between two specific states is given by summing the flip energies. As we average over all states, these energies are effectively independent, and $ \braket{\Delta E} \propto \braket{\delta_{nm}} $. For symmetry reasons, we find $ \braket{\delta_{nm}} = 0 $. Next, we turn to the variance. Summing over all possible states, we find the expression

\begin{align}
\mathrm{Var}(\Delta E) \propto 4 \tilde J^2 \bigg[\frac{1}{N} \sum_{n>1} \frac{1}{n^{2(1-\alpha)}} - \bigg( \frac{1}{N} \sum_{n>1} \frac{1}{n^{1-\alpha}} \bigg)^2 \bigg] \ .
\end{align}

At large $ N $, we can approximate the asymptotic behavior of these sums around $ \alpha \approx 0 $. We have $ \sum_{n>1} \frac{1}{n^{2(1-\alpha)}} \approx \pi^2 / 6 $. Additionally, we have $ \sum_n \frac{1}{n^{1-\alpha}} \approx \ln(N) $. This means we find

\begin{align}
\mathrm{Var}(\Delta E) \propto 4 \tilde J^2 \bigg[\frac{\pi^2}{6N} - \bigg( \frac{\ln(N)}{N} \bigg)^2 \bigg] \to 0 \ .
\end{align}

This means the variance approaches zero $ \propto 1/N $ to leading order with increasing system size. This is a direct consequence of the accumulation point in the single-particle spectrum. As the width of the multiplets vanishes, the associated timescale (the decay of the collective spins) must diverge. We highlight that this is a property of the spectrum not of any initial state. This explains why the lifetime of the collective spins does not depend on the fluctuations in the initial state but rather on the system size.

\section{Numerical Calculation of Microcanonical Expectation Values}

Here, we briefly describe the numerical method used to compute expectation values in the microcanonical ensemble, as presented in the main text (e.g., Figs.~1 and 3).

The microcanonical ensemble describes a system with a fixed energy $E$ (or within a narrow range $E$ to $E + \delta E$). We employ a random sampling approach to approximate expectation values in this ensemble. The method is outlined below, considering the specific form of our system's Hamiltonian in Eq.~(1) of the main text

\begin{align}
H = \frac{J}{N}(\boldsymbol{S}_A + \boldsymbol{S}_B) \cdot (\boldsymbol{S}_A + \boldsymbol{S}_B) + h (S_A^z - S_B^z) \ .
\end{align}
where $\boldsymbol{S}_A$ and $\boldsymbol{S}_B$ are the total spin vectors of sublattices $A$ and $B$ respectively, $N$ is the number of spins in each sublattice, $J$ is the coupling constant, and $h$ is the external magnetic field strength.

We begin by randomly sampling the orientations of the two spin sublattices, $A$ and $B$. Since the total spin of each sublattice, $\boldsymbol{S}_A^2$ and $\boldsymbol{S}_B^2$, are conserved quantities under this Hamiltonian, it suffices to sample their orientations independently. We generate random spherical angles $(\vartheta_A, \varphi_A)$ for sublattice $A$ and $(\vartheta_B, \varphi_B)$ for sublattice $B$, drawn from a uniform distribution over the sphere's surface. These angles define the orientations of the total spin vectors $\boldsymbol{S}_A$ and $\boldsymbol{S}_B$.

The energy $E$ of each configuration is then calculated using the given Hamiltonian.  We can express the Hamiltonian in terms of the sampled angles and system parameters as follows

\begin{align}
E(\vartheta_A, \varphi_A, \vartheta_B, \varphi_B) = \frac{J}{N} (2S(S+1) + 2S^2 (\sin\vartheta_A \sin\vartheta_B \cos(\varphi_A - \varphi_B) + \cos\vartheta_A \cos\vartheta_B)) + h\ S (\cos\vartheta_A - \cos\vartheta_B) \ .
\end{align}

where $S$ is the magnitude of the spin of a sublattice (which is fixed).

Subsequently, we check if this computed energy $E$ falls within the desired microcanonical window $[E_0, E_0 + \delta E]$. If the configuration is accepted ($E$ is within the window), the value of the observable of interest, $O(\vartheta_A, \varphi_A, \vartheta_B, \varphi_B)$, is calculated and added to a running sum.

These steps are repeated for $N_{\text{iter}}$ iterations. The microcanonical expectation value $\langle O \rangle_{mc}$ is then approximated as

\begin{align}
\langle O \rangle_{mc} \approx \frac{1}{N_{\text{accepted}}} \sum_{i=1}^{N_{\text{accepted}}} O(\vartheta_A^{(i)}, \varphi_A^{(i)}, \vartheta_B^{(i)}, \varphi_B^{(i)}) \ .
\end{align}

where $N_{\text{accepted}}$ is the number of accepted configurations and $(\vartheta_A^{(i)}, \varphi_A^{(i)}, \vartheta_B^{(i)}, \varphi_B^{(i)})$ is the $i$-th accepted configuration.

This method relies on the ergodicity assumption. The energy window width $\delta E$ should be chosen judiciously to balance accuracy and acceptance rate. The number of iterations $N_{\text{iter}}$ should be sufficiently large to ensure convergence of the expectation value. Proper algorithms must be employed for uniform sampling of points on a sphere to ensure the correct distribution of spin orientations. This specific implementation takes advantage of the conserved quantities and expresses the energy directly in terms of the sampled angles and the Hamiltonian's parameters $J$, $h$, $N$, and $S$.

\bibliography{references}